\documentclass[a4paper,10pt]{article}

\usepackage[T2A]{fontenc}

\usepackage[english]{babel}

\usepackage{graphicx}
\usepackage{amsthm, amsmath, amssymb, amsbsy, amsfonts, dsfont}
\usepackage{srcltx, color}

\newcommand{\PT}{\mathcal{PT}}

\def\tht{\theta}

\def\e{\varepsilon}
\def\g{\gamma}

\def\l{\lambda}
\def\p{\partial}

\def\d{\delta}
\def\L{\Lambda}

\def\vt{\vartheta}

\def\iu{\mathrm{i}}

\def\Op{\mathcal{H}}

\def\cL{\mathcal{L}}

\def\cH{\mathcal{H}}

\def\cG{\mathcal{G}}

\def\cI{\mathcal{I}}
\def\cJ{\mathcal{J}}
\def\cK{\mathcal{K}}

\def\Dom{\mathfrak{D}}

\newcommand{\R}{{\mathds{R}}}
\newcommand{\pxx}{{\partial_x^2}}
\newcommand{\pyy}{{\partial_y^2}}
\newcommand{\sech}{{\operatorname{sech}}}

\newcommand{\hg}{{\hat{g}}}
\newcommand{\heta}{{\hat{\eta}}}
\newcommand{\hU}{{\hat{U}}}
\newcommand{\bq}{{\textbf{q}}}
\def\bzeta{\boldsymbol{\zeta}} 
\newcommand{\bv}{{\textbf{v}}}

\DeclareMathOperator{\RE}{Re}
\DeclareMathOperator{\IM}{Im}

\newtheorem{theorem}{Theorem}[section]

\newcommand{\rev}[1]{\textcolor{black}{#1}}

\begin{document}

\allowdisplaybreaks

\title{Eigenvalues bifurcating from  the continuum  \\ in two-dimensional potentials generating \\ non-Hermitian gauge fields}

\author
{D. I. Borisov$^1$, D. A. Zezyulin$^2$\footnote{Corresponding author}}

\vskip -0.5 true cm

\maketitle

\begin{center}
	{\footnotesize $^1$
		Institute of Mathematics, Ufa Federal Research Center, Russian Academy of Sciences, Ufa, Russia,
		\\
		\&
		\\
		University of Hradec Kr\'alov\'e,  Hradec Kr\'alov\'e, Czech Republic
		\\
		{\tt borisovdi@yandex.ru}}
\end{center}

\begin{center}
	{\footnotesize $^2$
		School of Physics and Engineering, ITMO University, St. Petersburg 197101, Russia
		\\
		{\tt d.zezyulin@gmail.com}}
\end{center}

\begin{abstract}
	
It has been recently shown that complex two-dimensional (2D) potentials $V_\e(x,y)=V(y+\iu\e\eta(x))$ can be used to emulate non-Hermitian matrix gauge fields in optical waveguides. Here $x$ and $y$ are the   transverse coordinates, $V(y)$ and $\eta(x)$ are  real functions,    $\e>0$ is a small parameter, and $\iu$ is the imaginary unit. The real potential $V(y)$ is required to  have at least two discrete eigenvalues in the corresponding 1D Schr\"odinger operator. When both transverse directions are taken into account, these eigenvalues become \emph{thresholds} embedded in the continuous spectrum of the 2D operator.  Small nonzero  $\e$ corresponds to a non-Hermitian perturbation which can result in a bifurcation of  each threshold into an eigenvalue. Accurate analysis of these eigenvalues is important for understanding the behavior and stability of optical waves propagating in the artificial non-Hermitian gauge potential. Bifurcations of complex eigenvalues out of the continuum  is  the main object of the present  study. Using recent mathematical results from the rigorous analysis  of elliptic operators, we obtain  simple asymptotic expansions in $\e$ that describe   the behavior of bifurcating eigenvalues. The lowest threshold can bifurcate into a single eigenvalue, while every other threshold can bifurcate into a pair of complex eigenvalues. These   bifurcations   can be controlled by  the Fourier transform of function $\eta(x)$ evaluated at certain isolated points of the reciprocal space.  When the bifurcation does not occur,   the continuous spectrum of 2D operator contains a quasi-bound-state  which is   characterized by a strongly localized central peak coupled to small-amplitude but  nondecaying  tails. The analysis is applied to the case examples of parabolic and double-well potentials $V(y)$. In the latter case, the bifurcation of complex eigenvalues can be dampened if the two wells are widely separated.

\medskip


\end{abstract}

\section{Introduction}

Active interest to physics of non-Hermitian gauge potentials has been initiated by the   works of Hatano and Nelson who demonstrated that an imaginary vector potential can be used to control the localization-delocalization   in random systems \cite{Hatano1996,Hatano1997}.  
The understanding of this phenomenon has been further developed in a series of publications, see e.g. \cite{Brouwer,Efetov,Goldsheid,Hatano1998,Yurkevich,Takeda,CompGauge2D,Heinrichs}.  Most of these studies considered systems governed by non-Hermitian quantum mechanical Hamiltonians. In the meantime,  using  the well-known quantum-optical analogy (see \cite{analogy} for a review), it is possible to realize  imaginary vector potentials in modern photonics systems. In this context,   artificial non-Hermitian gauge fields have been proposed to facilitate   robust light transport in non-Hermitian  lattices \cite{LonghiGattiPRB,LonghiGattiSciRep,Longhi2017PRA,Qin,Zhang}. 
In  coupled slab waveguides, imaginary vector potentials can be used to enhance optical forces acting on  photons \cite{Lana}.

\begin{figure}
	\begin{center}
		\includegraphics[width=0.4\textwidth]{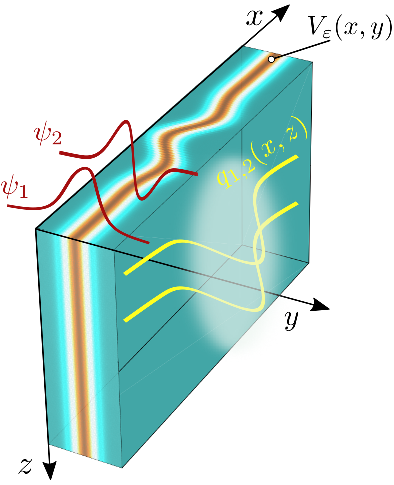}
	\end{center}
	\caption{Schematics of a three-dimensional non-Hermitian optical waveguide which contains a guiding slab with complex-valued dielectric  permittivity which is locally perturbed in the $x$-direction, \rev{and can be described using a complex-valued optical potential $V_\e(x,y)$}. Considered as a function of $y$, at each $x$ the resulting optical potential supports two (or more) guided modes $\psi_1$, $\psi_2$, \ldots. \rev{If the   field propagating along the $z$ axis is represented  in the form of a two-mode substitution (\ref{eq:twomodeintro}),} then the localized variation of the dielectric permittivity along the $x$-direction creates a matrix  non-Hermitian gauge potential  for envelopes $q_{1,2}(x,z)$, see Eq.~(\ref{eq:qgauge}).  	}
	\label{fig:scheme}
\end{figure}

Hermitian gauge fields are naturally present in description of various physical processes and, moreover, can be artificially synthesized   for cold atoms \cite{GaugeReview2, Galitski, GaugeReview},  microcavity exciton-polaritons \cite{polaritons},  as well as  in arrays of optical waveguides \cite{optics} and in electronic circuits \cite{circuits}.    However the implementation of non-Hermitian vector potentials remains  a much more challenging task. Most of presently  available  proposals are designed for light propagating in tight-binding discrete lattices, where the imaginary gauge field emerges as a result of the unbalanced hopping rates between the adjacent sites. In such tight-binding lattices, the effective gauge potential typically acts on a scalar field. At the same time,  it has been recently demonstrated \cite{Zez21} that a \emph{spatially continuous 
non-Hermitian matrix   gauge potential}  can be emulated in  an optical waveguide whose complex-valued dielectric permittivity varies in both  transverse directions  \rev{$(x,y)$ and is constant along the propagation direction ($z$), see schematics in Fig.~\ref{fig:scheme}.  It is assumed that  the   optical potential supports two guided modes, denoted as $\psi_1(x,y)$ and $\psi_2(x,y)$. Then the amplitude of the electric field $A(x,y,z)$ propagating in the waveguide is sought in the form of a two-mode substitution: 
\begin{equation}
\label{eq:twomodeintro}
A(x,y, z) = \psi_1(x,y)q_1(x,z) +  \psi_2(x,y)q_2(x,z),
\end{equation}
where $q_1(x,z)$ and $q_2(x,z)$ are the     envelopes. Substituting the  ansatz  (\ref{eq:twomodeintro}) in the paraxial Schr\"odinger-like equation,  it is possible to show that the propagation of the fields $q_1(x,z)$ and $q_2(x,z)$  along the $z$ direction can be described by means of  a pair of   equations    coupled by  a matrix $x$-dependent non-Hermitian gauge potential (their  detailed derivation will be presented below).} This obtained gauge-field system has been demonstrated in \cite{Zez21} to feature some intriguing behaviors which include   superexponential amplification and power blowup. Moreover, the optical realization   enables the account  of  nonlinear effects for   waves propagating in non-Hermitian gauge fields.

Regarding the specific choice of an optical  potential  which can be used for the    implementation of   non-Hermitian gauge fields, one of the promising candidates is given by     \emph{imaginary shifted} potentials of the form
\begin{equation}
\label{eq:Ve}
V_\e(x,y) = V(y + \iu\e\eta(x)),
\end{equation}
where $V(y)$ is real-valued potential, $\eta(x)$ is a bounded real-valued function, and $\e\geqslant 0$ is a parameter which governs the imaginary shift amplitude. If $\e$ is small enough, it is natural to expect that the eigenvalues of the one-dimensional (1D)  Schr\"odinger operator $-\p_y^2 + V(y + \iu\e\eta(x))$, where $x$ is treated as a parameter,  do not depend on $x$, while the eigenfunctions of this operator can be obtained from the eigenfunctions of   potential $V(y)$ by the   imaginary shift $y\mapsto y+\iu\e\eta(x)$.    This fact strongly  simplifies the construction of the gauge potential.

It is relevant to notice that if   $V(y)$ is an even function, then the imaginary-shifted potential $V_\e(x,y)$ given by equation  (\ref{eq:Ve}) is partially parity-time ($\PT$-) symmetric \cite{PPT}, i.e., obeys the following property: $V_\e^*(x,y) = V_\e(x,-y)$. It implies that complex eigenvalues of this potential exist as complex-conjugate pairs. We also mention   that similar imaginary-shifted potentials with $\eta(x)$ being constant have been considered earlier in the context of $\PT$-symmetric quantum mechanics and optics  \cite{Znojil,Bender,KSZ12,ZezKon12,Gallo}.

\rev{Regarding the applicability of the non-Hermitian gauge model,  we note that  the two-mode approximation (\ref{eq:twomodeintro})  does not provide a complete account of   2D  modes propagating  in the waveguide. Due to an inevitable error, the input beam $A(x,y, z=0)$ will never coincide exactly with the shape prescribed by the two-component substitution (\ref{eq:twomodeintro}). Therefore its propagation can excite additional guided modes which are not accounted by the spinor gauge-field model.  The complete spectrum of such modes is determined   by the 2D  Schr\"odinger operator
\begin{equation}\label{eq:He}
\Op_\e = -\pxx - \pyy + V_\e(x,y).
\end{equation}
Since the 2D   potential $V_\e(x,y)$ is complex-valued, operator (\ref{eq:He}) is non-Hermitian and therefore can have   complex eigenvalues which  correspond to unstable modes whose amplitudes  grow unbounded along the propagation distance. The instability increment is determined by the value of the    positive imaginary part of a complex eigenvalue. Thus an accurate analysis of complex eigenvalues of Schr\"odinger operator (\ref{eq:He})  becomes important for understanding the behavior of  linear waves and solitons guided by the  artificial non-Hemitian gauge potential.}  

\emph{The main goal of this paper} is the analysis   of complex eigenvalues emerging in 2D imaginary shifted potentials. Using   recent mathematical results from the rigorous  theory of elliptic operators, we argue that as $\e$ departs from zero, pairs of complex-conjugate eigenvalues can bifurcate from certain \emph{threshold points} of the continuous spectrum. Treating $\e$ as a small parameter, we construct asymptotic expansions for bifurcating eigenvalues and argue that in the generic situation their imaginary parts are of order  $\e^4$. Moreover, we show that the leading coefficient of the expansion for the  imaginary part  of   a  bifurcated   eigenvalue  can be made zero by a suitable  choice of function $\eta(x)$. In this case, the eigenvalue  behavior   is  described by the next orders of the perturbation theory. On the basis of a numerical study we  conjecture  that  in this case   the bifurcation from the continuum  can be suppressed completely, and, instead  of bound states with complex eigenvalues,   operator  $\cH_\e$  acquires   generalized eigenfunctions in its continuous spectrum whose amplitudes are strongly localized but nevertheless have small-amplitude  nondecaying tails. Such generalized eigenfunctions therefore  resemble   bound states in the continuum (BICs).

Besides of the importance  for engineering non-Hermitian gauge fields, in a more general context it should be   emphasized that the analysis of eigenvalues bifurcating from the continuum after a non-Hermitian perturbation is a separate direction of active research. In non-Hermitian potentials, the bifurcations of this type can be considered as a  mechanism of the phase transition from all-real to   complex spectrum \cite{Garmon,Yang17,KZ17} which is distinctively different from the  more familiar  exceptional-point  phase transition   that occurs   when  two (or more) discrete eigenvalues coalesce. In the 1D geometry,   bifurcation of a complex-conjugate pair  out of the continuum  usually occurs after the formation of the self-dual spectral singularity \cite{Most} in the continuous spectrum \cite{KZ17}, which corresponds to the waveguide operating in the  laser-absorber regime  \cite{Longhi}. Complex eigenvalues bifurcating from   interior points of the continuous spectrum are sometimes considered as a non-Hermitian generalization of BICs    \cite{BICoptics,Kartashov}. In this context,   our results offer   a mechanism for the controllable generation of such non-Hermitian BICs in  imaginary-shifted potentials.

The mathematical background of our study relies on a recent work published in Ref.~\cite{BZZ21} and  coauthored by the present authors. It furnishes     sufficient conditions for bifurcation of eigenvalues  out of the continuous spectrum  as a  multidimensional Hamiltonian is perturbed by a localized non-Hermitian perturbation.   The  mathematical result of \cite{BZZ21} is  very general and applies to  operators acting in       any dimensions  and   perturbations of rather general forms. However the application of this result to a specific problem can be sophisticated, which  justifies the relevance of present work. In addition, we note that the analysis presented herein has   some mathematical novelty: while the results of \cite{BZZ21} applied  only to Hamiltonians with bounded   potentials, here we extend the consideration onto some unbounded ones.

The rest of this paper is organized as follows. In Sec.~\ref{sec:gauge} we illustrate how the imaginary-shifted potentials can be used for generation of non-Hermitian gauge fields. Section~\ref{sec:problem} is dedicated to  a more  detailed formulation of the main objective of this paper which  is the study of eigenvalues bifurcating from the continuous spectrum of imaginary shifted potentials. In Sec.~\ref{sec:formulae} we present a general solution of the formulated problem, while in the two next sections we apply the results to case examples of parabolic (Section~\ref{sec:x2}) and double-well (Section~\ref{sec:double}) potentials. Section~\ref{sec:concl} concludes the paper.
Some auxiliary details and calculations have been moved to Appendices~\ref{sec:AppC}--\ref{sec:AppB}.

\section{ \rev{Model of the matrix  gauge potential, its basic properties, and applicability}}
\label{sec:gauge}

In this section, we show how the imaginary shifted potential (\ref{eq:Ve}) can be used to  emulate a non-Hermitian matrix potential in an optical waveguide. The construction develops from      the ideas sketched in \cite{Zez21}.

Let us consider   propagation of a paraxial light  beam along the $z$ axis. We assume that complex-valued refractive index of the guiding medium is described by function $V_\e(x,y)$ defined in (\ref{eq:Ve}). The propagation in  the longitudinal $z$  direction  obeys  the paraxial equation whose structure is equivalent to the 2D  Schr\"odinger equation:
\begin{equation}
\label{eq:optics}
\iu  \partial_z A = \cH_\e  A.
\end{equation}
Herein  $A = A(x,y,z)$ is the dimensionless  amplitude of the electric field, and operator $\cH_\e$ has been introduced  in Eq.~(\ref{eq:He}).

Let the real potential $V(y)$, which corresponds to zero imaginary shift, i.e., to $\e=0$ in equation (\ref{eq:Ve}), have two discrete eigenstates:
\begin{equation*}
\big(-\p_y^2 + V(y)\big)\psi_i(y)=\Lambda_i\psi_i(y)\quad\text{for}\quad y\in\R,\qquad i=1,2,
\end{equation*}
where the eigenfunctions $\psi_i(y)$, $i=1,2$, are real-valued,   normalized, and mutually orthogonal in $L_2(\R)$, and $\Lambda_i$ are the corresponding real eigenvalues. We further assume that the potential $V$ is bounded from below and its derivative does not grow faster than the potential. We also suppose that the potential admits an analytic continuation into the   strip  $\{y+\iu h:\ |h|<h_0\}$, where $h_0$ is some fixed sufficiently small number, and  the analytic continuation does not increase the growth. Rigorously these conditions are formulated as
\begin{equation}\label{2.0}
V(y)\geqslant C_1,\qquad |V'(y)|\leqslant C_2 |V(y)|+C_3,
 \qquad |V(y+\iu h)|\leqslant C_4 |V(y)|+C_5,\qquad y\in \mathds{R}, \quad |h|<h_0,
\end{equation}
where $C_i$, $i=1,\ldots,5$, are some constants; the constant $C_1$ may be negative, while other constants are positive. Our conditions for the potential $V$ guarantee the following facts, which will be proved in Appendix~\ref{sec:AppD}. The eigenfunctions $\psi_i$ can be also analytically continued into the complex plane, namely, into a strip $\{y+\iu h:\ |h|<h_1\}$, where $h_1$ is some sufficiently small number and $h_1<h_0$. These eigenfunctions solve the eigenvalue equation
\begin{equation}\label{2.1}
\big(-\p_y^2 + V(y+\iu h)\big)\psi_i(y+\iu h)=\Lambda_i\psi_i(y+\iu h)\quad\text{for}\quad y\in\R,\qquad i=1,2,
\end{equation}
with the same eigenvalues $\L_i$, which turn out to be independent of $h$. The eigenfunctions $\psi_i(y+\iu h)$ are the elements of the Sobolev space $W_2^1(\R)$ for each $h$, they are holomorphic in $h$ in the sense of the norm of this space and in particular this means that
\begin{equation}\label{2.4}
\int\limits_{\R} \big(|\psi_i(y+\iu h)|^2+|\psi_i'(y+\iu h)|^2\big)\,dy\leqslant c_1,
\end{equation}
where $c_1$ is some fixed constant independent of $h\in(-h_1,h_1)$.  At the same time, the eigenfunctions $\psi_i(y+\iu h)$ satisfy additional 
conditions:
\begin{equation}\label{2.2}
\begin{gathered}
\int\limits_{\R} \psi_i^2(y+\iu h)\,dy=1,\qquad \int\limits_{\R} \psi_i(y+\iu h)\psi_i'(y+\iu h)\,dy=0,\qquad i=1,2,
\\
\int\limits_{\R} \psi_1(y+\iu h) \psi_2(y+\iu h)\,dy=0,
\end{gathered}
\end{equation}
for $h\in(-h_1,h_1)$. Since
\begin{equation*}
0=\int\limits_{\R} (\psi_1(y)\psi_2(y))'\,dy=\int\limits_{\R} \psi_1(y)\psi_2'(y)\,dy + \int\limits_{\R} \psi_1'(y)\psi_2(y)\,dy,
\end{equation*}
we can introduce constant $\cI_0$ defined as follows:
\begin{equation*}
\cI_0:=\int\limits_{\R} \psi_1'(y)\psi_2(y)\,dy= - \int\limits_{\R} \psi_1(y)\psi_2'(y)\,dy.
\end{equation*}
Constant $\cI_0$ is real and we assume that it is non-zero. By the Cauchy integral theorem    we  conclude that
\begin{equation}\label{2.3}
\int\limits_{\R} \psi_1'(y+\iu h)\psi_2(y+\iu h)\,dy=\cI_0,\qquad   \int\limits_{\R} \psi_1(y+\iu h)\psi_2'(y+\iu h)\,dy=-\cI_0,\qquad h\in(-h_1,h_1).
\end{equation}

Let us look for   solution of equation (\ref{eq:optics}) in the form of the following two-mode substitution:
\begin{equation}
\label{eq:twomode}
 A(x,y,z) = \psi_1(y+\iu\e \eta(x)) q_1(x,z)  + \psi_2(y+\iu\e \eta(x)) q_2(x,z),
\end{equation}
where $q_i(x,z)$, $i=1,2$, are  envelopes. We assume that $\e \eta(x)$ is sufficiently small, namely, $\e |\eta(x)| < h_1$. Using   substitution (\ref{eq:twomode}) in paraxial equation (\ref{eq:optics}), multiplying the result by $\psi_1(y+\iu\e\eta(x))$ and $\psi_2(y+\iu\e\eta(x))$, and integrating over $y\in \R$, we obtain a system of two coupled equations that describe    evolution of the envelopes $q_i(x,z)$, $i=1,2$:
\begin{equation}
\begin{aligned}
&\iu \partial_z q_1= -\p_x^2q_1 + 2\iu\e\eta'(x)\cI_0\p_xq_2 + \big(\e^2 ( \eta'(x))^2(\cJ_1 - \Lambda_1) + \Lambda_1\big) q_1 + \big(\e^2( \eta'(x))^2\cJ + \iu\e\eta''(x) \cI_0\big)q_2,
\\
&\iu \partial_z q_2 = -\p_x^2q_2 - 2\iu\e\eta'(x)\cI_0\p_xq_1 + \big(\e^2 ( \eta'(x))^2(\cJ_2 - \Lambda_2) + \Lambda_2\big) q_2 + \big(\e^2( \eta'(x))^2\cJ - \iu\e\eta''(x) \cI_0\big)q_1,
\end{aligned}\label{eq:long1}
\end{equation}
where, for the sake of brevity,  the following coefficients have been introduced:
\begin{equation*}
\cJ_i = \int\limits_{\R} V(y) \psi_i^2(y)\,dy,\qquad \cJ = \int\limits_\R V(y)\psi_1(y)\psi_2(y)\,dy.
\end{equation*}
We note  that if $V(y)$ is an even function,   then eigenfunctions $\psi_1(y)$ and $\psi_2(y)$ associated with the subsequent eigenvalues $\Lambda_1$ and $\Lambda_2$ are of opposite parity and therefore $\cJ=0$.

We introduce  a non-Hermitian matrix     operator of the form
\begin{equation}
\label{eq:Pi}
\Pi := -\iu\p_x \sigma_0 - \iu\e \cI_0  \eta'(x) \sigma_2,
\end{equation}
where $\sigma_0$ is the 2$\times$2 identity matrix, and $\sigma_2$ is the second Pauli matrix:
\begin{equation}
\sigma_0 =  \begin{pmatrix}
1 & 0\\
0 & 1
\end{pmatrix},\quad \sigma_2 =  \begin{pmatrix}
0 & -\iu\\
\iu & 0
\end{pmatrix}.
\end{equation}
The term $ - \iu\e \cI_0 \eta'(x) \sigma_2$ can be interpreted as a non-Hermitian matrix gauge  potential. Then system (\ref{eq:long1}) can be rewritten in the following  compact  form
\begin{equation}
\label{eq:qgauge}
\iu \partial_z \bq  = \Pi^2 \bq + \mathrm{M} \bq,
\end{equation}
where $\bq = (q_1, q_2)^T$, and $ \mathrm{M} $ is an additional  Hermitian  matrix potential
\begin{equation}
\mathrm{M} =  \begin{pmatrix}
\e^2 (\eta'(x))^2 \big(\cI_0^2  + \cJ_1 - \Lambda_{1}\big) + \Lambda_1 &\iu\e\eta''(x)\cI_0 + \e^2(\eta'(x))^2 \cJ
\\[2mm]
-\iu\e\eta''(x)\cI_0 + \e^2(\eta'(x))^2 \cJ &\e^2 (\eta'(x))^2 \big(\cI_0^2 + \cJ_2 - \Lambda_{2}\big) + \Lambda_2.
\end{pmatrix}.
\end{equation}

\rev{
Model (\ref{eq:qgauge}), initially introduced in \cite{Zez21}, describes the evolution of a 1D spinor   $\bq(x,z)$ under the action of the non-Hermitian gauge field given by equation (\ref{eq:Pi}). It should be emphasized that equation~(\ref{eq:qgauge}) actually represents the minimal (i.e., the simplest) model. Its properties can be enriched by taking into    account    additional potentials and nonlinear effects.}

\rev{Minimal model  (\ref{eq:qgauge}) can be analyzed conveniently using the $z$-independent  carrier states   $\bzeta_{1,2}(x)$ which are introduced as a pair of mutually orthogonal  solutions to the following equations:
\begin{equation}
\Pi \bzeta_1(x) = 0, \qquad \Pi\bzeta_2(x) = 0, \qquad \bzeta_1^\dag\bzeta_2 = \bzeta_2^\dag\bzeta_1 = 0.
\end{equation}
For the gauge field given by Eq.~(\ref{eq:Pi}), we can choose
\begin{equation}
\bzeta_1(x) = e^{\e\cI_0\eta(x)} \left(
\begin{array}{c}
1\\-\iu
\end{array}
\right), \qquad \bzeta_2(x) = e^{-\e\cI_0\eta(x)} \left(
\begin{array}{c}
1\\ \iu
\end{array}
\right).
\end{equation} 
Then, representing the spinor field as a linear combination of carrier states,
\begin{equation}
\bq(x,z) = v_1(x, z) \bzeta_1(x) + v_2(x, z) \bzeta_2(x), 
\end{equation}
where $v_1(x,z)$ and $v_2(x,z)$ are new functions, from Eq.~(\ref{eq:qgauge}) we obtain 
\begin{equation}
\label{eq:vgauge}
\iu \partial_z \bv  =  -\partial_x^2 \bv +   \mathrm{U}(x) \bv,
\end{equation}
where  $\bv = (v_1, v_2)^T$, and $\mathrm{U}(x)$ is a $2\times 2$ matrix function whose entries are given as $\mathrm{U}_{i,j}(x) = \bzeta_i^\dag \mathrm{ M} \bzeta_j$, $i=1,2$, $j=1,2$. Matrix   $\mathrm{U}(x)$ is Hermitian, and therefore the field $\bv(x,z)$ conserves its   $L_2(\mathbb{R})$-norm:
\begin{equation}
\frac{d\ }{dz} \int_\mathbb{R} (|v_1|^2 + |v_2|^2) dx  = 0.
\end{equation}
Returning to the field $\bq(x,z)$, we readily compute
\begin{equation}
|q_1(x,z)|^2 + |q_2(x,z)|^2  = 2|v_1(x,z)|^2 e^{2\e \cI_0\eta(x)}  +  2|v_2(x,z)|^2 e^{-2\e \cI_0\eta(x)}.
\end{equation}
Therefore, for any bounded function $\eta(x)$ the $L_2(\mathbb{R})$-norm  of   $\bq(x,z)$ also remains globally bounded for any propagation distance:
\begin{equation}
\label{eq:bound}
\int_\mathbb{R}  (|q_1(x,z)|^2 + |q_2(x,z)|^2)  \   dx < Q_0 \quad \mbox{\quad for all \quad $z$},
\end{equation}
where $Q_0$ is a $z$-independent constant  determined by the initial conditions. Reconstructing the amplitude $A(x,y, z)$ from  the two-mode approximation  (\ref{eq:twomode}), we observe that inequality (\ref{eq:bound}) implies that the optical energy flow $\int_{\mathbb{R}^2} |A(x,y,z)|^2dx\,dy$ is also bounded. }

\rev{
Let us now discuss the applicability of the non-Hermitian gauge-field model  (\ref{eq:qgauge}). It has been obtained with   the two-mode substitution (\ref{eq:twomode}) and therefore is fully applicable only to optical beams of the specific shape. If the input beam $A(x, y,z=0)$ is not perfectly prepared, its propagation will deviate from the prediction of   gauge-field system (\ref{eq:qgauge}).  A more complete description of the beam propagation can be obtained using   the   2D Shr\"odinger operator $\cH_\e$ which enters the paraxial equation (\ref{eq:optics}).  Since the   2D Shr\"odinger operator with the complex-valued potential $V_\e(x,y)$ is not self-adjoint, its spectrum,  generically speaking,    contains complex eigenvalues whose impact can be  dramatic. The energy flow  of optical guided modes   associated with  complex eigenvalues   can  grow indefinitely,  in contrast to the behavior predicted by the two-component gauge-field model where   the   energy flow  $\int_{\mathbb{R}^2} |A(x,y,z)|^2dx\,dy$ is bounded. Therefore, the     unstable (i.e., growing along the propagation distance) eigenmodes can result in  a significant discrepancy between the predictions of the non-Hermitian gauge field system and the actual propagation of the beam in the 2D waveguide, and   an accurate analysis of    eigenvalues of the 2D Schr\"odinger operator becomes   relevant.}

\section{Formulation of the main problem}
\label{sec:problem}

\begin{figure}
	\begin{center}
		\includegraphics[width=0.75\textwidth]{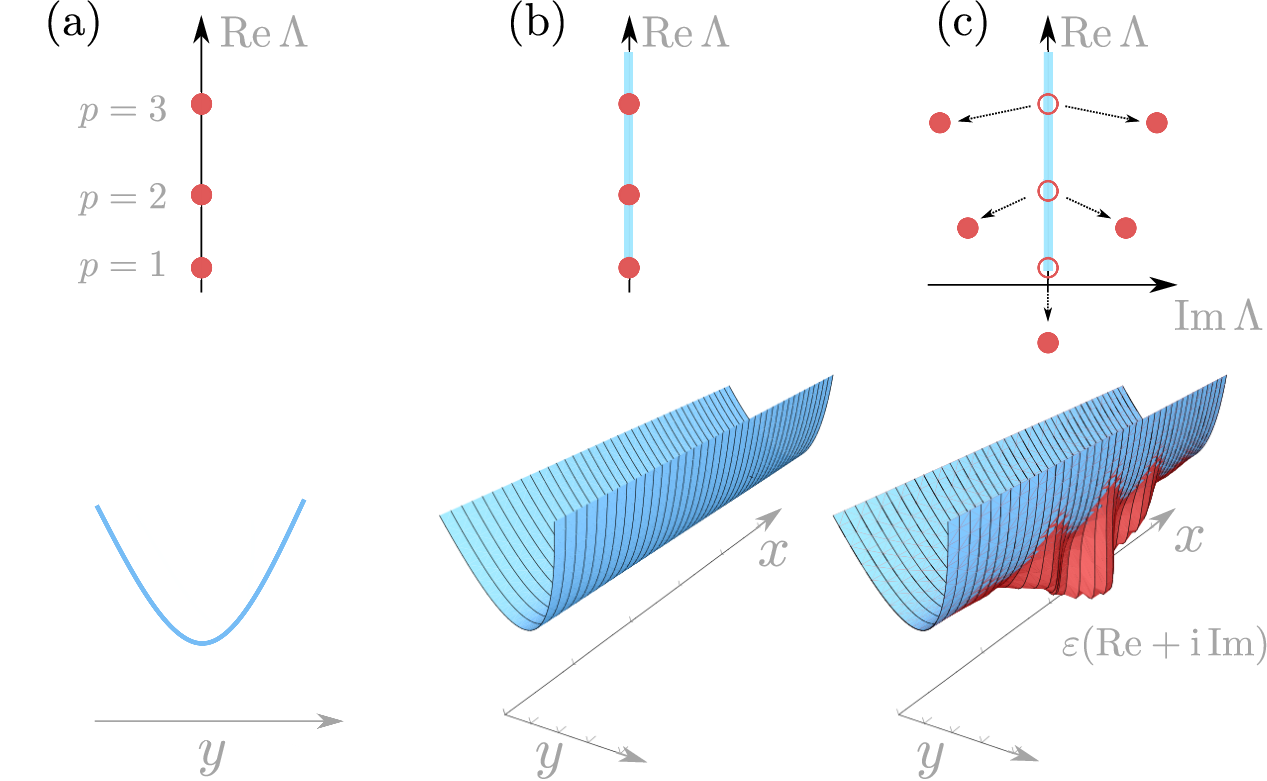}
	\end{center}
	\caption{(a) One-dimensional real potential $V(y)$ with a (finite or infinite) sequence of discrete eigenvalues $\Lambda_1<\Lambda_2<\ldots<\Lambda_{p}<\ldots$ (b) When the second dimension (i.e., the $x$ axis) is added, the potential $V(y)$ becomes a cylindrical surface, and  the former discrete eigenvalues become thresholds embedded in the continuous spectrum of 2D   states. (c) When a localized complex-valued   perturbation \rev{[schematically shown as a red surface next to the writing  $\e(\RE +\iu \IM)$]}  is added to the  potential, the internal thresholds   bifurcate out of the continuum in  2D eigenvalues  associated to  $L_2(\R^2)$-integrable eigenfunctions, and the lowest (edge) threshold bifurcates into a single eigenvalue.}
	\label{fig:01}
\end{figure}

In this section, we proceed to the main question set in this paper, which is the behavior of the spectrum of the 2D operator $\cH_\e$ given by Eq.~(\ref{eq:He}) with the imaginary shifted potential $V_\e(x,y)$ given by  Eq.~(\ref{eq:Ve}), for small values of the perturbation parameter $\e$. As required by  construction of the gauge potential outlined in Sec.~\ref{sec:gauge}, we assume that the 1D operator  $\cL_0 = -\pyy + V(y)$, where $V(y)$ is a
real potential, has  at least  two  discrete eigenvalues below its essential spectrum (which may  be empty).  All these eigenvalues are simple and therefore can be arranged in the ascending order: $\Lambda_1<\Lambda_2<\ldots$.  Let the corresponding eigenfunctions be denoted as $\psi_1(y)$, $\psi_2(y)$, etc. These eigenfunctions are real-valued and orthonormalized in $L_2(\R)$. Considering now the 2D operator $\cH_\e$ given by (\ref{eq:He}),   numbers $\Lambda_1, \Lambda_2, \ldots$ do not   correspond to its eigenvalues.   Instead, they become \emph{thresholds}   embedded in the continuous spectrum associated with     2D spatially extended   modes,  see the  illustration in Fig.~\ref{fig:01}. More precisely, the lowest 1D  eigenvalue $\Lambda_1$ is situated at the bottom of the continuous spectrum of the 2D operator. We therefore call $\Lambda_1$ the \emph{edge threshold}. Higher eigenvalues $\Lambda_2<\Lambda_3<\ldots$  are situated at    internal points of the continuum. We therefore call them \emph{internal thresholds}. When the     potential is perturbed with a 2D non-Hermitian perturbation, which in our case results from the imaginary shift with localized function $\eta(x)$, the thresholds can generate eigenvalues of the 2D operator which bifurcate from the continuous spectrum (as schematically shown in Fig.~\ref{fig:01}). Eigenfunctions corresponding to the bifurcating eigenvalues are square integrable in $\R^2$.   A detailed study of bifurcations of these eigenvalues is the main objective of our study.

The mathematical background of our work relies on   recent results from the rigorous theory of elliptic operators presented in \cite{BZZ21}. While the  mathematical result of \cite{BZZ21}      concerns   operators acting in any dimensions  and applies to  perturbations of rather general forms, 
its adaptation to the specific problem can be sophisticated. We therefore find it appropriate to present an adjusted  formulation of this result in Appendix~\ref{sec:AppA}, where two theorems are formulated: one is for bifurcation from the edge threshold, and another one is for internal thresholds. These theorems also include   asymptotic expansions which characterize the location of   bifurcating eigenvalues for small $\e$.

All conditions for the potential $V$ needed for our analysis have already been formulated in the previous section. Regarding the constraints on the function $\eta(x)$ which determines the spatial shape of the imaginary shift,  in this work
we consider the case when $\eta(x)$ is piecewise continuous function
decaying at least exponentially at infinity. Namely, we assume that this function satisfies the estimate
\begin{equation}\label{eq:eta}
|\eta(x)|\leqslant C e^{-a|x|}\quad\text{for all}\quad x\in\R,
\end{equation}
where $C$ and $a$ are some positive constants independent of $x$.  Notice that the construction of gauge field outlined in Sec.~\ref{sec:gauge} additionally requires that $\eta(x)$ is twice differentiable.

\section{ General formulae for bifurcating eigenvalues}
\label{sec:formulae}

To examine   how each threshold $\L_p$, $p=1,2,\ldots$,  bifurcates from the continuous spectrum into   an eigenvalue (or eigenvalues) of  operator $\cH_\e$, we are going to apply Theorem~\ref{thm:p=1} (for the edge threshold) and Theorem~\ref{thm:p>1} (for the internal thresholds) from Appendix~\ref{sec:AppA}. To  enable  these theorems, we first represent the operator $\cH_\e$ in form (\ref{A3}) by expanding the potential $V_\e(x,y)$ in a series with respect to small $\e$. By the Taylor  theorem   with the  Lagrange form of the remainder we  get:
\begin{equation}
\label{eq:Vexp}
V_\e(x,y) = V(y) + \iu\e \eta(x) V'(y)  -  \frac{\e^2}{2}\eta^2(x)  V''(y)
 -\frac{\iu\e^3}{6}\eta^3(x)V'''\big(y+\iu\tht(x,y)\big),
\end{equation}
where $\tht(x,y)$ is   some real-valued function, such that  $|\theta(x,y)|\leqslant \e |\eta(x)|$.
Then we denote
\begin{equation}
\label{eq:V1V2}
V_1(x,y):=\iu\eta(x)V'(y),\qquad V_2(x,y):=-\frac{\eta^2(x)}{2}V''(y), \qquad
V_3(x,y):=-\frac{\iu\eta^3(x)}{6}V'''\big(y+\iu\tht(x,y)\big).
\end{equation}
We shall show, see inequalities (\ref{8.2}) in Appendix~\ref{sec:AppC}, that conditions (\ref{2.0})
imposed on the potential $V(y)$ imply inequalities (\ref{A2}), and therefore the theorems do apply.

Further, for each $p$ we introduce the following constants which correspond to the coefficients of asymptotic expansions for    eigenvalues bifurcating from the threshold $\Lambda_{p}$:
\begin{align*}
&K_{1,p}=-\frac{1}{2}\int\limits_{\R^2} \psi_p^2(y) V_1(x,y) \,dxd y,
\\
&K_{2,p,\tau}= -\frac{1}{2}\int\limits_{\R^2}\left(\psi_p^2(y) V_2(x,y) - \psi_p(y) V_1(x,y)\cG_{p,\tau}( V_1 \psi_p)\right)\,dxdy,
\end{align*}
where  $\cG_{p,\tau}$ {is a linear operator} defined by Eq.~(\ref{eq:Gp}), and $\tau \in \{-1, +1\}$ is a parameter.
Sufficient conditions for   threshold $\Lambda_p$ to  bifurcate into an eigenvalue (or two eigenvalues)   depend on    real and imaginary  parts of the constants $K_{1,p}$ and $K_{2,p,\tau}$. However, irrespectively of the particular  form of potential $V(y)$,  the leading-order correction $K_{1,p}$  is zero for each $p$, in view of the identity
\begin{equation}
\label{eq:zero}
\int\limits_{\R} V'(y)  \psi_p^2(y)\,d y=0,
\end{equation}
which can be checked easily by integrating by parts and using the eigenvalue equation for $\psi_p$. This result should be regarded as a special feature of potential  $V_\e(x,y)$ that results specifically from the  imaginary shift.

Proceeding to the next coefficient $K_{2,p,\tau}$, we note that for the edge threshold (i.e., for $p=1$) this  coefficient 
is in fact independent of $\tau$, i.e., $K_{2,1,+1} = K_{2,1,-1} =: K_{2,1}$. According to Theorem~\ref{thm:p=1}, the sufficient condition for the edge threshold to   bifurcate  into an eigenvalue  reduces to $\RE K_{2,1}>0$. If this condition is satisfied, then the edge threshold bifurcates into a single eigenvalue denoted as $\lambda_1$, and the corresponding asymptotic expansion has the form
\begin{equation}
\label{eq:asp=1}
\l_1(\e)=\L_1- \e^4 K_{2,1}^2+O(\e^5).
\end{equation}

Regarding  each internal threshold with $p\geqslant 2$, according to Theorem~\ref{thm:p>1}, the sufficient condition for the bifurcation to occur   becomes   $\RE K_{2,p,\tau}>0$  and $\tau \IM K_{2,1,\tau}>0$ for some $\tau\in \{+1, -1\}$.  {The emerging eigenvalue has the asymptotic expansion
\begin{equation}
\label{eq:asp>1}
\l_{p,\tau}(\e)=\L_p- \e^4 K_{2,p,\tau}^2+O(\e^5).
\end{equation}
If the above conditions for the real and imaginary parts of $K_{2,p,\tau}$ are satisfied for both $\tau=1$ and $\tau=-1$, then the internal threshold $\Lambda_{p}$ bifurcates in two different eigenvalues, each having asymptotic behavior as in Eq.~(\ref{eq:asp>1}).}

Let us now compute coefficients $K_{2,p,\tau}$, $p=1,2,\ldots$. For $V_1(x,y)$ and $V_2(x,y)$ given in Eq.~(\ref{eq:V1V2}), we can rewrite the formula for $K_{2,p,\tau}$ as
\begin{equation}\label{K2a}
K_{2,p,\tau}=\frac{1}{2}\int\limits_{\R^2} \left( \frac{\eta^2(x)}{2}  V''(y)   {\psi_p^2(y)} - \eta(x) V'(y)   \psi_p(y) U_{p,\tau}(x,y)\right)\,dxdy,
\end{equation}
where $U_{p,\tau} :=  \cG_{p,\tau}\left(  \eta(x) V'(y)  \psi_p(y)\right)$. If the potential $V(y)$ and  its  eigenfunction $\psi_p(y)$ are known, then  $U_{p,\tau}$ is  the only function in (\ref{K2a}) that needs to  be found. The structure of this function is given in Appendix~\ref{sec:AppA}  by     expression (\ref{eq:Gp}), where   one has to substitute   $f(x,y)= \eta(x) V'(y)  \psi_p(y)$. While, at first glance,  the resulting expression seems rather bulky, it can be in fact simplified significantly. Relegating technical calculations to Appendix~\ref{sec:AppB}, here we present the final result which conveys rather transparent information on the bifurcating eigenvalues.   Specifically,  for the real part of   coefficient $K_{2,p,\tau}$ we get
\begin{equation}
\label{eq:real}
\RE K_{2,p,\tau}=\frac{1}{2}\int\limits_{\R^2}\left(\frac{\eta^2(x)}{2}V''(y)\psi_p^2(y) -\eta(x)V'(y)\psi_p(y)U_p^\bot(x,y)\right)\,dxdy + \sum_{j=1}^{p-1}\frac{\cI_{pj}^2\cH_{p,j}}{4\sqrt{\Lambda_p-\Lambda_j}},
\end{equation}
where we have introduced new notations
\begin{align}
\label{eq:cI}
&\cI_{p,j}:= \int\limits_{\R} \psi_p(y)\psi_j(y)V'(y) \,d y,
\\
\label{eq:cH}
&\cH_{p,j}:=\pi^{-1}\,  \mathrm{v.p.} \int\limits_\R \frac{|\heta(s)|^2}{\sqrt{\Lambda_p-\Lambda_j}-s} \,d s.
\end{align}
Hereafter   we  use a hat symbol to denote the  Fourier transform which for any suitably behaved function $g(x)$ is defined as   $\hat{g}(s) =    \int\limits_{\R} g(x) e^{isx} \,d x$ with $s$ being the variable in the reciprocal space. Therefore, $\heta(s)$ is the Fourier transform of $\eta(x)$, and  $\cH_{p,j} $  is in fact the  Hilbert transform \cite{Brychkov} of function $|\heta(s)|^2$ evaluated at the point $\sqrt{\Lambda_p-\Lambda_j}$.   {The function} $U_p^\bot(x,y)$ in Eq.~(\ref{eq:real}) is purely real-valued and can be found as a solution of the inhomogeneous equation obeying certain orthogonality conditions, see Eqs.~(\ref{eq:Ubot}),~(\ref{eq:OrtCon}) in Appendix~\ref{sec:AppA}:
\begin{equation}\label{eq:Ubot:eta}
\big(-\pxx - \pyy +V(y) - \Lambda_p\big)U_p^\bot = \eta(x) \left(V'(y)\psi_{p}(y)  - \sum_{j=1}^{p-1} \cI_{p,j} \psi_j(y)\right).
\end{equation}
Notice that the upper limit in the latter sum is $p-1$ and not $p$, because from (\ref{eq:zero}) we have $\cI_{p,p}=0$.

For the imaginary part of   $K_{2,p,\tau}$ we compute  (see Appendix~\ref{sec:AppB} for details):
\begin{align}
\label{eq:imag}
\tau\IM K_{2,p,\tau}=-&\sum_{j=1}^{p-1}\frac{\cI_{pj}^2}{4\sqrt{\Lambda_p-\Lambda_j}} \big|\heta(\sqrt{\Lambda_p-\Lambda_j})\big|^2.
\end{align}

Equations (\ref{eq:real}) and, especially, (\ref{eq:imag}) are the central result of this paper. The following comments can be made.
\begin{enumerate}
	\item By definition, for the edge threshold we have $p=1$, and hence  the sums in (\ref{eq:real}) and (\ref{eq:imag}) disappear, and the corresponding coefficient $ K_{2,1,\tau}$ does not depend on $\tau$ and is purely real.
	According Theorem~\ref{thm:p=1} in Appendix~\ref{sec:AppA}, if $K_{2,1,-1}=K_{2,1,+1}>0$, then the edge threshold $\L_1$ bifurcates into a single simple eigenvalue $\l_1(\e)$ with the asymptotic behavior (\ref{eq:asp=1}).

	\item For the internal thresholds with $p\geqslant 2$, the real part $\RE K_{2,p,\tau}$ is independent on $\tau$. Most interestingly, the imaginary part can be controlled by the Fourier image $\heta(s)$ evaluated at certain isolated points of the reciprocal space  $s$. If for some fixed $p\geqslant 2$, the Fourier image  is not zero at some $s=\sqrt{\Lambda_p-\Lambda_j}$, then  we immediately have
	\begin{equation}\label{eq:ImtK2}
	\tau\IM K_{2,p,\tau}<0,
	\end{equation}
	and, if additionally   $\RE K_{2,p,\tau}>0$, then    Theorem~\ref{thm:p>1}  guarantees a pair of eigenvalues bifurcating from the internal threshold $\Lambda_p$. These two eigenvalues have asymptotic behavior written down in (\ref{eq:asp>1}). 	On the other hand, if for some $p$ the Fourier image is zero at each $s=\sqrt{\Lambda_p-\Lambda_j}$ with $j=1,\ldots, p-1$,  then the sufficient  conditions of  bifurcation do not  hold.  In this case the behavior of   internal threshold $\Lambda_p$ has to be found from next orders of the perturbation theory, which is beyond the scope of the present study.

	\item  The leading corrections in asymptotic expansions (\ref{eq:asp=1})--(\ref{eq:asp>1}) are of order $O(\e^4)$. This is in contrast to the   generic  situation treated in \cite{BZZ21}, where the leading corrections  were of order $O(\e^2)$. This fact is a peculiarity   of the imaginary shifted-potentials and follows from  identity (\ref{eq:zero}).
	
	\item   Function  $U_p^\bot$ in Eq.~(\ref{eq:real}) is not always available analytically, but  it can be  computed  numerically. It can be more efficient to employ Fourier transform in $x$, i.e., to compute the Fourier image  $\hU_2^\bot(s,y)$ from    equation
	\begin{equation*}
	(s^2-\pyy  + V(y) - \Lambda_p) \hU_p^\bot = 
\heta(s)\Bigg(V'(y)   \psi_p(y) - \sum_{j=1}^{p-1} \cI_{p,j} \psi_j(y)\Bigg),
	\end{equation*}
	which can be solved for each $s$;  the orthogonality condition
		\begin{equation*}
		\int\limits_{\R} \hU_p^\bot(s, y)\psi_j(y)\,d y=0,\qquad j=1,2, \ldots, p,
		\end{equation*}
		should be also satisfied. Once  the Fourier image $\hU_p^\bot(s,y)$ is obtained,  it is not necessary to invert the Fourier transform, because with the Parseval's equality the unknown integral in (\ref{eq:real}) can be computed as
	\begin{equation}
	\int\limits_{\R^2} \eta(x)V'(y)\psi_p(y)U_p^\bot(x,y)\,d x \,d y = (2\pi)^{-1} \int\limits_{\R^2} V'(y)\psi_p(y)\heta^*(s)\hU_p^\bot(s,y)\,d s\,d y,
	\end{equation}
where the superscript $^*$ denotes the complex conjugation.
	
\end{enumerate}

\section{Parabolic potential}
\label{sec:x2}
\subsection{Analytical results}

\textcolor{black}{In this section,  we consider}  an important particular case of the imaginary-shifted parabolic potential . Therefore in this section we  have $V(y)=y^2$, and    $\psi_{p}(y)$ are Gauss-Hermite functions:
\begin{equation}
\psi_p(y) = \frac{H_{p-1}(y)}{\sqrt{2^{p-1}(p-1)!\sqrt{\pi}}}e^{-y^2/2}, \quad p=1,2,\ldots,
\end{equation}
where $H_{p-1}(y)$ are Hermite polynomials. The thresholds are situated at $\Lambda_p=2p-1$.  For the Gauss-Hermite functions the following relations hold
\begin{equation}
\label{eq:GHprop}
\int\limits_\R y \psi_{p+1}(y)\psi_p(y)\,d y = \sqrt{\frac{p}{2}}, \qquad  \int\limits_\R y \psi_{p}(y)\psi_j(y)\,d y = 0 \quad \text{as}\quad |j-p|\geqslant 2,
\end{equation}
which in terms of the integrals $\cI_{p,j}$ introduced  in (\ref{eq:cI}) imply   that for each fixed $p$ we have $\cI_{p,j}=0$ except for
\begin{equation}
 \cI_{p,p-1} = \sqrt{2(p-1)}, \qquad \cI_{p,p+1} = \sqrt{2p}.
\end{equation}
Using these generalized orthogonality relations and solving Eq.~(\ref{eq:Ubot:eta}) with the separation of variables, for the edge threshold (i.e., for $p=1$)   we find  $U_1^\bot(x,y) = 
\textcolor{black}{c(x) \psi_2(y)}$, where
\begin{equation}
\label{eq:c(x)}
c(x) :=  \frac{1}{2}\int\limits_{\R}\eta(x_1)e^{-\sqrt{2}|x-x_1|}\,d x_1.
\end{equation}
Therefore, for $p=1$ the \textcolor{black}{constant $K_{2,1}$} 
acquires the form  (recall that for the edge threshold this coefficient does not depend on $\tau$):
\begin{equation}
\label{eq:edgex2}
K_{2,1} =\frac{1}{2}\int\limits_{\R}   \big(\eta^2(x)-  \sqrt{2} c(x)\eta(x)\big)\,dx.
\end{equation}
Using    the convolution theorem (which states that the Fourier  transform of the convolution  is equal to the  product of two Fourier transforms), for   Fourier transform of $c(x)$  we compute $\hat{c}(s) = {\sqrt{2}}{(s^2+2)^{-1}}\hat{\eta}(s)$. Together with the Parseval's equality this simplifies   Eq.~(\ref{eq:edgex2}) to
\begin{equation}
K_{2,1}=\frac{1}{4\pi}\int\limits_{\R}  \frac{s^2|\hat{\eta}(s)|^2}{s^2+2}\,d s.  
\end{equation}
This coefficient is positive, which means that for any $\eta(x)$   an  eigenvalue bifurcates  from the edge threshold.

For internal thresholds with $p\geqslant 2$, from    (\ref{eq:Ubot:eta})  we find    $U_p^\bot(x,y) = 
 \textcolor{black}{\sqrt{p} c(x) \psi_{p+1}(y)}$, and, after simplification, from (\ref{eq:real}) and (\ref{eq:imag}) we obtain
\begin{align}
\label{eq:kpx2Re}
&\RE K_{2,p,\tau}  =  \frac{1}{4\pi}\int\limits_{\R} |\hat{\eta}(s)|^2 \frac{s^2+2-2p}{s^2+2}\,ds  + \frac{p-1}{2\sqrt{2}\pi}\mathrm{v.p.}\int\limits_{\R} \frac{|\heta(s)|^2}{\sqrt{2}-s}\,ds,
\\
&\label{eq:kpx2Im}
\tau\IM K_{2,p,\tau} = -\frac{p-1}{2\sqrt{2}}|\heta(\sqrt{2})|^2.
\end{align}
The imaginary parts of  obtained coefficients are nonzero, except for the situation when
the  Fourier transform   $\heta(s)$ vanishes exactly  at  $s=  \sqrt{2}$.

\subsection{Numerical examples}

If $\eta(x)$ is given, then the calculation of coefficient $K_{2,p,\tau}$ reduces to evaluation of several integrals, whose analytical values are not always available, but the numerical evaluation is simple. Choosing as a model example $\eta(x) = \sech(\sqrt{2} x)$ and, respectively, $|\heta(s)|^2 = (\pi^2/2)\sech^2(\pi s\sqrt{2}/4)$, for three lowest thresholds we compute
\textcolor{black}{
\begin{align*}
&p=1:\qquad K_{2,1}=\frac{\sqrt{2}}{2} \left(1 - \frac{\pi^2}{12}\right)\approx 0.126,
\\
&p=2:\qquad K_{2,2,\pm}\approx 0.375 \mp \iu\,0.277,
\\
&p=3:\qquad K_{2,3,\pm}\approx 0.625 \mp \iu\,0.554,
\end{align*}
where $\approx$ corresponds to the numerical accuracy of the integration.  Therefore, the edge threshold bifurcates into a single simple eigenvalue with the asymptotic expansion $\l_1(\e)=1-\e^4 K_{2,1}^2+O(\e^5)$, while    internal thresholds with $p=2,3$  bifurcate into a pair of
eigenvalues $\l_{p,\pm }(\e) = 2p-1 - \e^4 K_{2,p,\pm}^2+O(\e^5)$.} Since in the case at hand the potential $V_\e(x,y)$ is (partially) $\PT$ symmetric, the eigenvalue $\lambda_1(\e)$  is real, and the   eigenvalues bifurcating from each threshold  form  a  complex-conjugate pair: $\l_{p, +}(\e) = \l_{p, -}^*(\e)$.

\begin{figure}
	\begin{center}
		\includegraphics[width=0.99\textwidth]{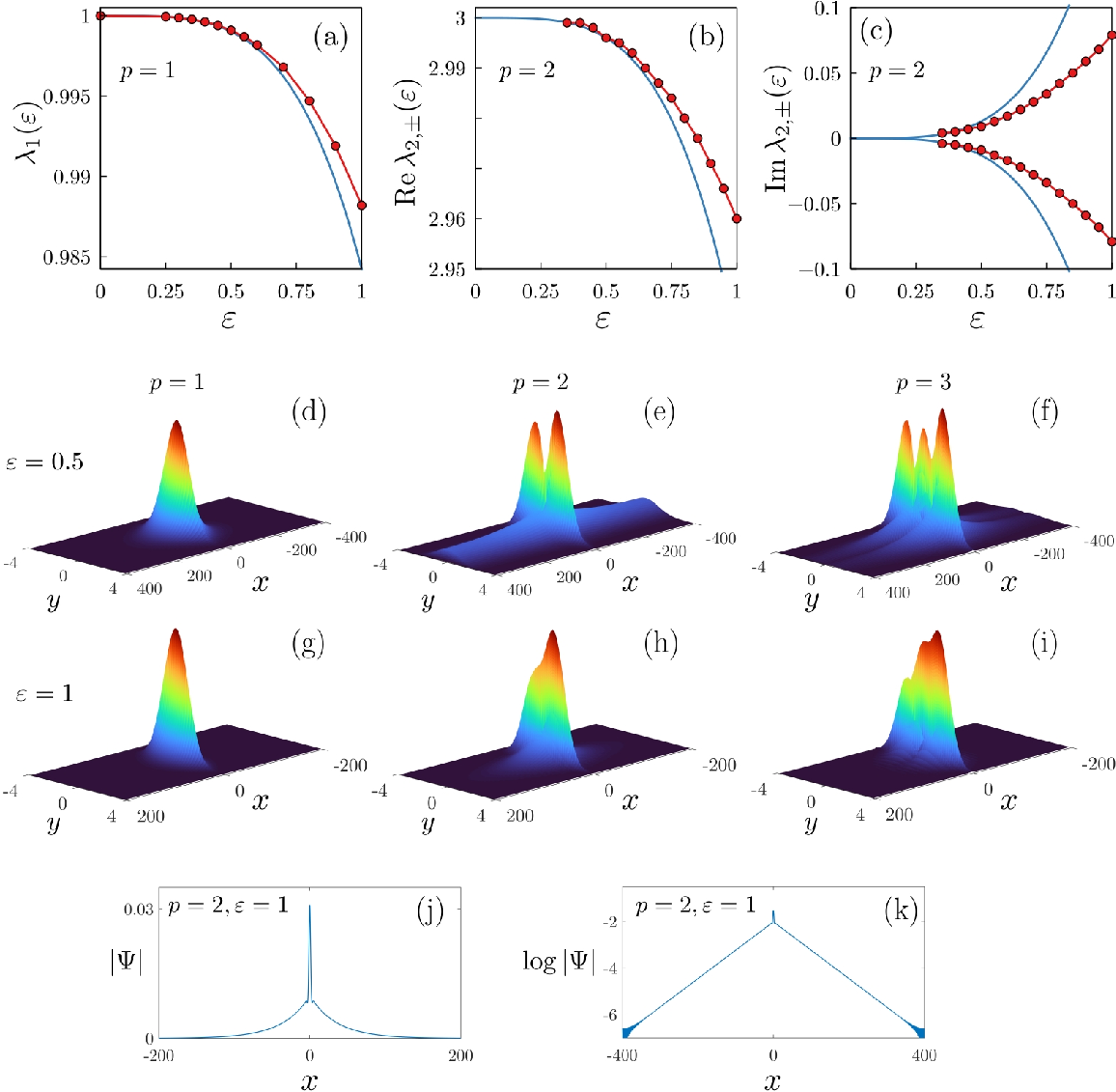}
	\end{center}
	\caption{(a--c) Analytical predictions for bifurcating eigenvalues (blue curves) and numerically computed eigenvalues (red circles connected with red lines). (a)  Real eigenvalue  $\lambda_1(\e)$ bifurcating from the edge threshold. (b,c) Real and imaginary parts of the complex-conjugate eigenvalues $\lambda_{2,\pm}(\e)$ bifurcating from the internal threshold $\Lambda_2$.  Panels (d--i) show moduli of bifurcating eigenfunctions at $\e=0.5$ (upper row)  and $\e=1$ (bottom row) for $p=1,2,3$. Two lowest panels (j,k) show the modulus (linear and log scales) of the   eigenfunction with $p=2$ and $\e=1$ plotted as a function of $x$ at fixed $y=0$. Throughout  this  figure  $\eta(x) = \sech(\sqrt{2} x)$}
	\label{fig:02}
\end{figure}

To confirm the  bifurcations of eigenvalues out of the continuum,  we have computed the    spectrum of 2D operator $\cH_\e$ using a finite-difference scheme which reduces the problem to finding   eigenvalues of a large sparse matrix. It should be emphasized that for small $\e$ (approximately for $\e\lesssim 0.5$) the accuracy of our numerics is strongly  limited by the extremely poor localization of bifurcating eigenfunctions along  the $x$ axis. This is especially true for   eigenvalues bifurcating from the internal thresholds, because the   tails  of the corresponding   eigenfunctions decay very slowly and oscillate. To detect these poorly localized eigenfunctions,   instead of the zero Dirichlet boundary conditions, one needs to use von Neuman  boundary conditions.  Nevertheless, the numerical results   confirm bifurcations of two-dimensional eigenfunctions   and exhibit  reasonable qualitative agreement with asymptotic predictions, see Fig.~\ref{fig:02}. As $\e$ increases, the     localization of bifurcating eigenstates enhances dramatically, which becomes evident  if one compares the  $x$-axes limits for eigenfunctions at $\e=0.5$ and $\e=1$ in Fig.~\ref{fig:02}(d--i).

\begin{figure}
	\begin{center}
		\includegraphics[width=0.99\textwidth]{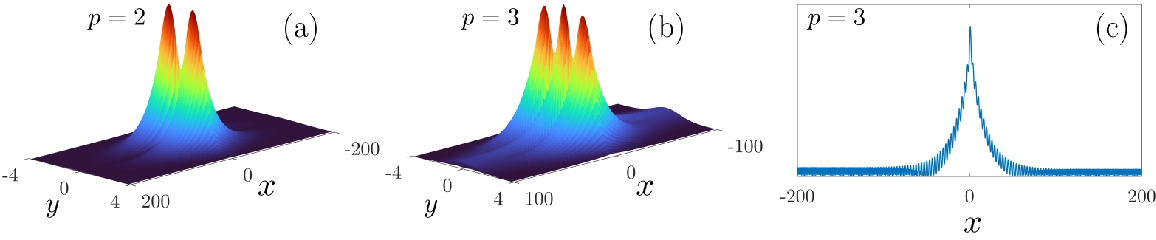}
	\end{center}
	\caption{(a,b) Moduli of quasi-BIC states  that emerge  in the vicinity of the internal thresholds with $p=2$ and $p=3$ when bifurcations of complex-conjugate eigenvalues do not occur. Here $\eta(x) =  e^{-x^2/2} + e^{-(x-\pi/\sqrt{2})^2/2}$ and $\e=0.25$. 		Panel (c) shows the modulus of the eigenfunction with $p=3$ plotted as a function of $x$ at fixed $y=0$. }
	\label{fig:03}
\end{figure}

According  to Eq.~(\ref{eq:kpx2Im}), the imaginary parts of bifurcating eigenvalues are  governed by the  Fourier  transform  $\heta(s)$  evaluated at only a single point $s=\sqrt{2}$. If the   Fourier transform vanishes at this point, then the bifurcations of internal thresholds are   described   by the next orders of the perturbation theory or possibly  even disappear   completely.  To enable this  situation,  we  {consider} another example of function $\eta(x)$ in the form   $\eta(x) = e^{-x^2/2} + e^{-(x-\pi/\sqrt{2})^2/2}$. For its Fourier transform we compute  $|\hat{\eta}(s)|^2 = 4\pi e^{-s^2}(1+\cos(s\pi/\sqrt{2}))$ and hence  $|\hat{\eta}(\sqrt{2})|^2=0$.  Computing numerically the corresponding spectrum  we cannot reliably detect any eigenvalue bifurcating from the internal thresholds. This allows us  to conjecture that in this case the bifurcations of complex eigenvalues  are suppressed completely. In the meantime,   in the vicinity of each internal threshold, we numerically observe      unusual states  in the continuous spectrum    which feature strong localization around the origin, see Fig.~\ref{fig:03}(a,b). However, a closer inspection reveals that such states do not represent genuine bound states in the continuum (BICs) but have   small-amplitude   tails that oscillate and  do not decay as $x$ approaches $\pm \infty$, as emphasized in  Fig.~\ref{fig:03}(c).

\begin{figure}
	\begin{center}
		\includegraphics[width=0.999\textwidth]{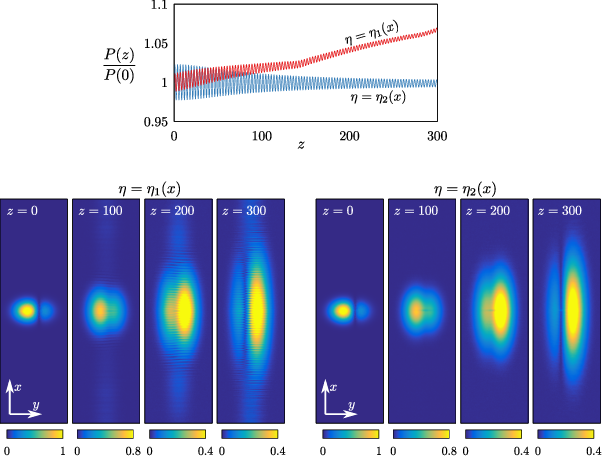}
	\end{center}
	\caption{\rev{Upper panel shows the energy flow $P(z) = \int_{\mathbb{R}^2} |A(x,y,z)|^2dxdy$ for two propagations computed from   Shr\"odinger equation (\ref{eq:optics}) with the imaginary-shifted potential $V_\e(x,y)$ obtained for      $ \eta_1(x) = e^{-x^2/2}$ and   $\eta_2(x) = e^{-x^2/2} + e^{-(x-\pi/\sqrt{2})^2/2}$. For each propagation, we used $\e=0.25$, and  the input beam $A(x,y, z=0)$ has been computed using the  two-mode substitution (\ref{eq:twomode}) with $q_1(x,z=0) = q_2(x, z=0) = e^{-(x/15)^2}$;   small random perturbation has been also added to the initial condition. Lower panels show pseudocolor snapshots of intensity $|A(x,y,z)|$ at different propagation distances $z$. Each of these plots  is shown in the   window $(x,y)\in [-4, 4]\times [-200, 200]$.  } }
	\label{fig:dynamics}
\end{figure}

\rev{We have additionally performed direct numerical simulation of the beam propagating in the Shr\"odinger equation (\ref{eq:optics}) with the imaginary-shifted parabolic potential. The input beam  $A(x,y, z=0)$ has been prepared in the form of   two-mode substitution (\ref{eq:twomode}) with
\begin{equation}
	  q_1(x,z=0) = q_2(x, z=0) = e^{-(x/15)^2}.
\end{equation}	
This initial condition has been additionally perturbed by a random noise whose maximal amplitude has been about 10$\%$ from the   amplitude of the unperturbed functions. The propagation has been computed for two different imaginary shifts   $\eta(x)$, namely $\eta_1(x) = e^{-x^2/2}$ and  $\eta(x) = e^{-x^2/2} + e^{-(x-\pi/\sqrt{2})^2/2}$.  The asymptotic expansions indicate that for $\eta=\eta_1(x)$ unstable eigenvalues are present in the spectrum, while for $\eta=\eta_2(x)$ the equality  $|\hat{\eta}(\sqrt{2})|^2=0$ takes place, which means that the   leading coefficients of asymptotic expansions  for   the imaginary parts of unstable eigenvalues  vanish, i.e., the instability is expected to be suppressed. In agreement with these expectations, for $\eta=\eta_1(x)$ we observe a distinctive growth of the energy flow   $P(z) = \int_{\mathbb{R}^2} |A(x,y,z)|^2dxdy$, while for $\eta=\eta_2(x)$ the    growth is suppressed.  The difference between the two   imaginary shifts  can be also observed in  the pseudocolor intensity snapshots    shown in the lower panels of Fig.~\ref{fig:dynamics}. For $\eta=\eta_1(x)$ the diffraction of the input beam is superimposed by the   noise growing with the propagation distance, while for $\eta = \eta_2(x)$ we observe a  much more clear diffraction picture, where the perturbation growth is suppressed. }

\section{Double-well potential}
\label{sec:double}

In this section, we apply results of our analysis to an example of a double-well potential  $V(y)$. Here we limit our exposition to only the first internal threshold  (i.e., that corresponding to $p=2$) which is the most important for the emulation of the non-Hermitian gauge field  (see Sec.~\ref{sec:gauge}). Equation (\ref{eq:imag}) allows to rewrite
\begin{align}
\label{eq:imag:dw}
\tau\IM K_{2,2,\tau}=-& \frac{\cI_{2,1}^2}{4\sqrt{\Lambda_2-\Lambda_1}} |\heta(\sqrt{\Lambda_2-\Lambda_1})|^2,
\end{align}
where, as defined in  (\ref{eq:cI}), $\cI_{2,1} = \int\limits_{\R}  V'(y)\psi_1(y)\psi_2(y)\,d y$.   Integrating by parts, we can rewrite
\begin{equation}
\cI_{2,1} = -\int\limits_{\R} V(y) (\psi_1(y)\psi_2(y))'\,d y = -(\Lambda_2-\Lambda_1)\int\limits_{\R} \psi_1'(y)\psi_2(y)\,d y.
\end{equation}

For a double-well potential $V(y)$, let us consider the asymptotic limit of widely separated wells (WSW-limit).  In this limit   we have $\Lambda_2-\Lambda_1\to 0$ and  $\int\limits_{\R} \psi_1'(y)\psi_2(y)\,d y \to 0$. Therefore, in the WSW-limit  we can  estimate $\cI_{2,1} = o(\Lambda_2-\Lambda_1)$. Hence, irrespectively of the choice of function $\eta(x)$, from (\ref{eq:imag:dw}) we have
\begin{equation}
\label{eq:Imk}
\tau \IM K_{2,p,\pm}  =o((\Lambda_2-\Lambda_1)^{\frac{3}{2}}) \quad\text{as}\quad  \Lambda_2-\Lambda_1\to 0.
\end{equation}
Suppose that a pair of eigenvalues bifurcate from the internal threshold $\Lambda_{2}$. According to our expansions, the imaginary parts of these    eigenvalues  can be approximated as $|\IM\lambda_{2,\pm}| \approx 2 \e^4 |\RE  K_{2,p,\pm}\IM K_{2,p,\pm}|$.  Even though the real part   $\RE  K_{2,p,\pm}$ is not yet computed explicitly (it is given by Eq.~(\ref{eq:real}) with $p=2$), it is rather natural to conjecture that   this quantity remains   finite in the WSW-limit, i.e., $\RE  K_{2,p,\pm} =  O(1)$. This conjecture immediately implies that  when  the distance between potential wells increases,  the imaginary parts of   bifurcating values decrease at fixed value of parameter $\e$.
 
\begin{figure}
	\begin{center}
		\includegraphics[width=0.65\textwidth]{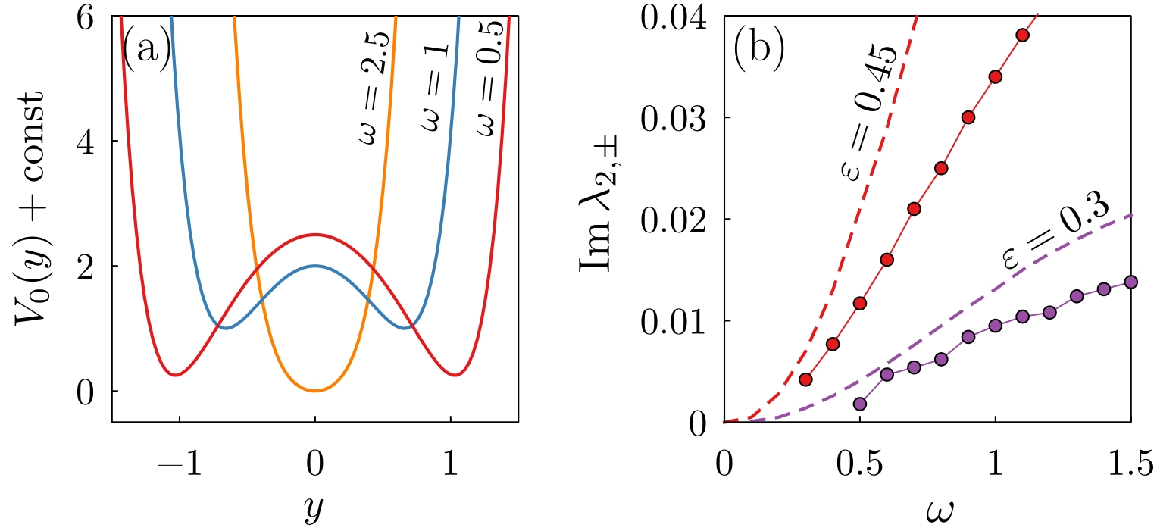}
	\end{center}
	\caption{(a) Plot of     potential in Eq.~(\ref{eq:razavy}) for $\omega=2.5$ (single-well) and $\omega=1$ and $0.5$ (double-well).  For better comparison, each  plot is  shown modulo to an additive  constant which shifts the curve  in the vertical axis. (b) Positive imaginary parts of bifurcating eigenvalues as functions of $\omega$ for fixed $\e=0.3$ and $\e=0.45$.  Dashed lines correspond to the predictions of asymptotic expansions, and circles connected with solid lines are numerical results. }
	\label{fig:razavy}
\end{figure}

To illustrate more explicitly  the  suppression    of  imaginary parts of  bifurcating eigenvalues,   we consider an example in the form of an exactly solvable potential     \cite{razavy1980}:
\begin{equation}
\label{eq:razavy}
V(y) = \omega^2\sinh^2(2y)  - 4\omega\cosh(2y),
\end{equation}
where $\omega>0$ is a positive parameter.   For   $\omega\geqslant 2$,  potential $V(y)$ is single-well with the global minimum  situated at $y=0$. The    value $\omega=2$ corresponds to a saddle-node bifurcation, and  for $\omega<2$ instead of the   central minimum the potential $V(y)$  has a local maximum at $y=0$,   while two new lateral  minima emerge at $y=\pm \frac{1}{2}\mathrm{acosh}(2\omega^{-1})$, see Fig.~\ref{fig:razavy}(a). Therefore, for this potential the WSW-limit corresponds to $\omega\to +0$.

The two  lowest eigenstates of this potential are  known explicitly:
\begin{align*}
&\psi_1(y) =  \sqrt{\frac{{2}}{ {\cK_1(\omega)+ \cK_0(\omega)}}}\exp\left(-\frac{\omega}{2}\cosh(2y)\right)\cosh(y),
\\
&\psi_2(y) = \sqrt{\frac{{2}}{ {\cK_1(\omega)- \cK_0(\omega)}}}\exp\left(-\frac{\omega}{2}\cosh(2y)\right)\sinh(y),
\end{align*}
where $\cK_{0,1}(\omega)$ are the modified Bessel functions. The corresponding eigenvalues read $\Lambda_{1,2} = \mp2\omega-1$. Therefore the meaning of parameter $\omega$ is rather transparent: it measures the distance between the eigenvalues, i.e., $\Lambda_2-\Lambda_1=4\omega$.

Taking advantage of the solvability of this double-well potential, we compute
\begin{equation*}
\cI_{2,1} = \frac{4\omega\cK_0(\omega)}{\sqrt{\cK_1^2(\omega)-\cK_0^2(\omega)}},
\qquad
\int\limits_\R V''(y)\psi_2^2(y)\,d y =  8\omega^2 +\frac{ 16\cK_1(\omega)}{\cK_1(\omega)-\cK_0(\omega)}.
\end{equation*}
Using a handbook asymptotic expansion  for  Bessel functions, for small positive $\omega$ we approximate $\cI_{2,1}\sim -4\omega^2\ln\omega$, which agrees with the  behavior  $\cI_{2,1} = o(\Lambda_2-\Lambda_1) = o(\omega)$  predicted  above  in the limit of small distance between the two eigenvalues. Computing numerically function $\hU^\bot_2(s,y)$,   for several particular values of $\omega$ we obtain
\begin{align*}
&\omega=0.1:\qquad K_{2,2,\pm}\approx  0.47 \mp \iu\, 7\cdot 10^{-3},
\\
&\omega=0.4:\qquad K_{2,2,\pm }\approx  0.81 \mp \iu\,0.18,\\
&\omega=1:\hphantom{.0}\qquad  K_{2,2,\pm }\approx  3.14 \mp \iu\,0.91.
\end{align*}
Thus the product of real and imaginary parts of the  coefficients $  K_{2,2,\pm }$ decreases with $\omega$, which decreases the imaginary part of bifurcating eigenvalues. This trend has been confirmed in our numerics, as illustrated in Fig.~\ref{fig:razavy}(b) which shows the dependence of $\IM\lambda_{2,\pm}$ on $\omega$ for two fixed valued of $\e$.

\section{Conclusion}
\label{sec:concl}

The motivation of this work has been twofold. First, it has been encouraged by the recent proposal on the implementation of matrix non-Hermitian gauge potentials in optical waveguides. The   derivation of the gauge potential   relies  on the two-mode substitution and does not  take into account the complete spectrum of 2D guided modes which are    naturally expected to  impact the  light  beam propagation in the  waveguide. In the present work, we have explicitly demonstrated that  non-Hermitian imaginary-shifted  optical potential can have complex eigenvalues which bifurcate from certain threshold points of the continuous spectrum. We have derived asymptotic expansions which can be used to  estimate the location of complex eigenvalues.  \rev{From the practical point of view, our results can be used to estimate the instability increment of guided modes  that emerge due to the inevitable difference between the two-mode approximation and a real-world signal propagating in the waveguide. We demonstrate that the unstable eigenvalues can be controlled and suppressed   by tuning the shape of the  imaginary shift. A numerical study confirms    that in certain situations  the bifurcations of unstable complex eigenvalues can be suppressed, and, instead of square-integrable bound states,  the continuous spectrum of the  perturbed potential contains quasi-BIC states.}

From a more general point of view, our work has been motivated by the growing   interest to bound states that bifurcate out of the continuous spectrum of certain operators subject to non-Hermitian perturbations. \rev{Since  the real parts of the bifurcating eigenvalues lie inside the interval  of propagation constants occupied by  the continuous spectrum, the corresponding modes can be interpreted as non-Hermitian counterparts of BIC states. Therefore our results offer    a mechanism for a systematic generation  of such non-Hermitian BIC states.  Using our approach it is possible to create multiple pairs of    complex-conjugate non-Hermitian BICs   that bifurcate from different threshold points  of the continuous spectrum.}

Finally, we note that  similar eigenvalue bifurcations can be a reason of weak instabilities of two-dimensional solitons existing near an exceptional point in the   $\PT$-symmetric waveguide \cite{EPsolitons}. Indeed, while the   one-dimensional solitons obtained from the two-mode approximation are stable, they become metastable (i.e., weakly unstable) when the second transverse dimension is taken into account. An accurate analysis of this issue is a relevant task for future work. Another subject that deserves  further study is the generalization of the present analysis from localized imaginary shift to a periodic one,  i.e., from exponentially decaying to periodic functions $\eta(x)$.

\section*{CRediT authorship contribution statement}

All authors contributed equally to this work.

\section*{Declaration of competing interest}

The authors declare that they have no known competing financial interests or personal relationships that could have appeared to influence the work reported in this paper.

\section*{Data availability}

No data was used for the research described in the article.

\section*{Acknowledgments}

The authors thank K.~Pankrashkin for a valuable discussion on the domains of operators with growing potentials. D.A.Z. is grateful to  Vladimir Konotop for useful discussions.

\section*{Funding}

The research presented in Sections~2,~3,~4 is financially supported by Russian Science Foundation, grant no. 23-11-00009.

\appendix

\section{Properties of potential}\label{sec:AppC}

Here we prove a few auxiliary inequalities for the potential $V$ and its derivatives. These inequalities will be then employed for proving the analyticity of the eigenfunctions in Appendix~\ref{sec:AppD}, which was mentioned in Section~\ref{sec:gauge},  and also they have been employed to adapt the general approach from \cite{BZZ21} to our case in Section~\ref{sec:formulae}.

A basic auxiliary inequality we are going to prove reads as
\begin{equation}\label{8.1}
C_6\Big(\max\limits_{t\in[y-\d,y+\d]}|V(t)|+1\Big)\leqslant |V(y)|+1 \leqslant \max\limits_{t\in[y-\d,y+\d]}|V(t)|+1,\qquad y\in\mathds{R},
\end{equation}
for some sufficiently small $\d$ and some positive constants $C_6$ and $C_7$ independent of $y$.  The upper bound is obvious and we need to confirm only the lower one. Indeed, for $y\in\mathds{R}$, $t\in[-\d,\d]$ with an arbitrary $\d>0$  we have an obvious formula
\begin{equation*}
V(y)=V(y+t)-\int\limits_{y}^{y+t} V'(\ell)\,d\ell,
\end{equation*}
which  by the second inequality in (\ref{2.0}) yields
\begin{equation*}
|V(y)|\geqslant |V(y+t)|-C_2\d \sup\limits_{\ell\in[y-\d,y+\d]} |V(\ell)|-C_3 \d.
\end{equation*}
Taking then the maximum over $t\in [-\d,\d]$, we find:
\begin{equation*}
|V(y)|\geqslant (1-C_2\d) \max\limits_{\ell\in[y-\d,y+\d]} |V(\ell)| - C_3\d.
\end{equation*}
Choosing   $\d$ small enough to ensure the inequalities $(1-C_2\d)>\frac{1}{2}$, $C_3\d<\frac{1}{2}$, we arrive at the lower bound in (\ref{8.1}).

Inequality (\ref{8.1}) allows us to establish important estimates for the derivatives of the function $V$, which have been employed in Section~\ref{sec:formulae}  for adapting general approach from \cite{BZZ21} to our case and also they are used for proving the analyticity of the eigenfunctions $\psi_i$ in Appendix~\ref{sec:AppD}. Namely, for an arbitrary point $z=y+\iu h$ with  $|h|<\frac{h_0}{2}$  by the Cauchy integral formula we have a formula for the $n$th derivative of the potential $V$:
\begin{equation*}
V^{(n)}(z) =\frac{n!}{2\pi\iu} \int\limits_{\g_z} \frac{V(\xi)\,d\xi}{(\xi-z)^{n+1}},\qquad \g_z:=\bigg\{\xi:\, |\xi-z|=\frac{h_0}{3}\bigg\}.
\end{equation*}
This formula, the third inequality in (\ref{2.0}) and (\ref{8.1}) yield:
\begin{equation}\label{8.2}
|V^{(n)}(y+\iu h)|\leqslant n!\left(\frac{3}{h_0}\right)^n \max\limits_{t\in[y-\d,y+\d]} |V(t)| \leqslant C_7(n) \big(|V(y)|+1\big),
\end{equation}
where $C_7(n)$ is some constant independent of $y$ and $h$.

We shall also need one more estimate for $V$, which will be used for establishing   the analyticity of the eigenfunctions $\psi_i$ in Appendix~\ref{sec:AppD}; this estimate reads as
\begin{equation}\label{8.3}
|V'(y)|\leqslant |V(y)|^\frac{3}{2}+C_8,\qquad y\in\mathds{R},
\end{equation}
with some positive constant  $C_8$  independent of $y$. If $|V(y)|\geqslant C_7^\frac{1}{3}(1)$, by (\ref{8.2}) we immediately get:
\begin{equation*}
|V'(y)|\leqslant C_7(1)(|V(y)|+1)\leqslant C_7(1)\frac{|V(y)|^\frac{4}{3}}{|V(y)|^\frac{1}{3}} + C_7(1)\leqslant |V(y)|^\frac{4}{3}+C_7(1).
\end{equation*}
For $|V(y)|<C_7^\frac{1}{3}(1)$, again by (\ref{8.2}) we obtain:
\begin{equation*}
|V'(y)|\leqslant C_7^\frac{4}{3}(1)+C_7(1)\leqslant |V(y)|^\frac{4}{3}+C_7^\frac{4}{3}(1)+C_7(1).
\end{equation*}
Two latter estimates prove (\ref{8.3}) with $C_8:=C_7^\frac{4}{3}(1)+C_7(1)$.

\section{Analyticity of eigenfunctions}\label{sec:AppD}

Here we establish the analyticity of the eigenfunctions and other properties stated in Section~\ref{sec:gauge}. In the space $L_2(\R)$ we define a self-adjoint operator with the differential expression
\begin{equation*}
\cL_0=-\frac{d^2\ }{dy^2}+V(y)
\end{equation*}
as associated with a corresponding closed symmetric sesqulinear form. According \cite[Eq. (1.2)]{Metafune}, \cite[Thm. 1]{EvGi}, inequality (\ref{8.3}) ensures that the domain of the operator $\cL_0$ is given by the identity
\begin{equation}\label{A4}
\Dom(\cL_0)=W_2^2(\R)\cap L_2\big(\R, (1+|V(y)|^2)\,dy\big),
\end{equation}
where $W_2^2(\R)$ is the Sobolev space (see e.g. \cite{Lieb}). We additionally assume that the operator  $\cL_0$ has discrete eigenvalues below the edge of its essential spectrum. These eigenvalues are denoted by $\L_1<\L_2<\ldots<\L_p<\ldots$, and   the associated orthonormalized in $L_2(\R)$ eigenfunctions are denoted by $\psi_p=\psi_p(y)$. The operator $\cL_0$ can have an empty essential spectrum and in such case we just deal with infinitely many eigenvalues $\L_p$.

Then we introduce one more operator in $L_2(\R)$ with the differential expression
\begin{equation*}
\cL_h=-\frac{d^2\ }{dy^2}+V(y+\iu h),\qquad |h|<\frac{h_0}{2},
\end{equation*}
on the same domain as $\cL_0$, see (\ref{A4}). The potential in this operator satisfies the third inequality in (\ref{2.0}) and this guarantees that the operator $\cL_h$ is well-defined and closed. Owing to the analyticity and estimates (\ref{8.2}), the resolvent of the operator $\cL_h$ is holomorphic in sufficiently small $h$.  Hence, the operator $\cL_h$ is holomorphic in the generalized sense, see \cite[Ch. V\!I\!I, Sec. 1.2]{Kato}. Then by \cite[Ch. V\!I\!I, Sec. 1.3, Thms.~1.7,~1.8]{Kato}, the operator $\cL_h$ possesses simple eigenvalues $\tilde{\L}_i=\tilde{\L}_i(h)$, $i=1,\ldots,p$, and associated eigenfunctions $\tilde{\psi}_i=\tilde{\psi}_i(y,h)$, $i=1,\ldots,p$, which are holomorphic in $h\in(-h_1,h_1)$ with some $h_1<\frac{h_0}{2}$ such that
\begin{equation}\label{8.9}
\tilde{\L}_i(0)=\L_i,\qquad \tilde{\psi}_i(y,0)=\psi_i(y),\qquad i=1,\ldots,p.
\end{equation}
The holomorphy of the eigenfunctions holds in the sense of $L_2(\R)$-norm and in view of the eigenvalues equations and (\ref{A4}) we then see that they are also holomorphic in the sense of the norm in $\Dom(\cL_0)$. Since the eigenfunctions $\psi_i(y)$ are ortonormalized, by the above holomorphy we conclude that
\begin{equation*}
\bigg|\int\limits_{\R} \tilde{\psi}_i^2(y {, h})\,dy\bigg|\geqslant C_9>0,\qquad i=1,\ldots,p,
\end{equation*}
provided $h_1$ is small enough,
where $C_9$ is some fixed constant independent of $h$. Then without loss of generality we can normalize the eigenfunctions $\tilde{\psi}_i$ as
\begin{equation}\label{8.5}
\int\limits_{\R} \tilde{\psi}_i^2(y,h)\,dy=1.
\end{equation}
Differentiating this identity in $h$, we get:
\begin{equation}\label{8.6}
\int\limits_{\R}\tilde{\psi}_i(y,h)\frac{\p\tilde{\psi}_i}{\p h}(y,h)\,dy=0.
\end{equation}
We also have
\begin{equation}\label{8.7}
\int\limits_{\R} \tilde{\psi}_i(y,h)\frac{\p\tilde{\psi}_i}{\p y}(y,h)\,dy=\frac{1}{2} \int\limits_{\R} \frac{\p\tilde{\psi}_i^2}{\p y}(y,h)\,dy=0,
\end{equation}
as well as an analogue of identity (\ref{eq:zero}):
\begin{equation}\label{8.8}
\int\limits_{\R} V'(y+\iu h)\tilde{\psi}_i(y,h)\frac{\p\tilde{\psi}_i}{\p y}(y,h)\,dy=0.
\end{equation}

We differentiate the eigenvalue equation for $\tilde{\psi}_i$ in $y$ and $h$ and this leads us to extra two equations:
\begin{equation}\label{8.10}
(\cL_h-\tilde{\L}_i)\frac{\p\tilde{\psi}_i}{\p h}=   \frac{\p \tilde{\L}_i}{\p h} \tilde{\psi}_i - \iu V' \tilde{\psi}_i,\qquad (\cL_h-\tilde{\L}_i)\frac{\p\tilde{\psi}_i}{\p y}=\tilde{\L}_i \frac{\p\tilde{\psi}_i}{\p y}  - V' \tilde{\psi}_i,\qquad V'=V'(y+\iu h).
\end{equation}
We multiply the first equation by $\tilde{\psi}_i$ and integrate over $\R$ using (\ref{8.5}), (\ref{8.7}), (\ref{8.8})  and the eigenvalue equation for $\tilde{\psi}_i$:
\begin{equation*}
\frac{\p\tilde{\L}_i}{\p h}=\int\limits_{\R} \tilde{\psi}_j (\cL_h-\tilde{\L}_i)\frac{\p\tilde{\psi}_i}{\p h}\,dy=\int\limits_{\R} \frac{\p\tilde{\psi}_i}{\p h} (\cL_h-\tilde{\L}_i)\tilde{\psi}_j\,dy=0,
\end{equation*}
and therefore,  due to (\ref{8.9}), the eigenvalues $\tilde{\L}_i$ are independent of $h$ and $\tilde{\L}_i(h)=\L_i$. Comparing then equations in (\ref{8.10}), we see that the functions $\iu\frac{\p\tilde{\psi}_i}{\p h}$ and $\frac{\p\tilde{\psi}_i}{\p y}$ can differ only by an eigenfunction $\tilde{\psi}_i$ multiplied by some constant. However, it follows from orthogonality conditions (\ref{8.6}), (\ref{8.7}) that such constant should vanish and hence,
\begin{equation*}
\iu\frac{\p\tilde{\psi}_i}{\p h}(y,h)=\frac{\p\tilde{\psi}_i}{\p y}(y,h).
\end{equation*}
This identity is in fact the Cauchy-Riemann condition  ensuring that the eigenfunctions $\tilde{\psi}_i(y,h)$ are holomorphic functions of the complex variable $y+\iu h$. Redenoting them by $\psi_i(y+\iu h)$, we conclude that the eigenfunctions $\psi_i=\psi_i(y)$ of the operator $\cL_0$ admit an analytic continuation into the strip $\{y+\iu h:\, |h|<h_1\}$.

Multiplying the eigenvalue equation for the function $\psi_i$ by $\psi_j$ and integrating twice by parts over $\R$, we get
\begin{align*}
\L_i\int\limits_{\R} \psi_i(y+\iu h)\psi_j(y+\iu h)\,dy=& \int\limits_{\R} \psi_j(y+\iu h)\cL_h \psi_i(y+\iu h)\,dy
\\
=&\int\limits_{\R} \psi_i(y+\iu h)\cL_h \psi_j(y+\iu h)\,dy=\L_j\int\limits_{\R} \psi_i(y+\iu h)\psi_j(y+\iu h)\,dy,
\end{align*}
and since the eigenvalues $\L_i$ and $\L_j$ do not coincide, we find:
\begin{equation}\label{8.4}
\int\limits_{\R} \psi_i(y+\iu h)\psi_j(y+\iu h)\,dy=0,\qquad i\ne j.
\end{equation}

The above established properties of the functions $\psi_i$ ensure all facts on these functions employed in Section~\ref{sec:gauge}.

\section{Theorems on bifurcating eigenvalues for two-dimensional potentials}
\label{sec:AppA}

Here we formulate an adaption of a rather     general result on eigenvalues bifurcating from   threshold points  in the essential spectra,   which has recently been  formulated and proven in \cite{BZZ21}.

By $V_1=V_1(x,y)$, $V_2=V_2(x,y)$, $V_3=V_3(x,y,\e)$ we denote three complex-valued piece-wise continuous potentials defined on $\R^2$, where $\e$ is a small positive parameter. We assume that these potentials satisfy the estimates
\begin{equation}\label{A2}
|V_1(x,y)|+|V_2(x,y)|+|V_3(x,y,\e)|\leqslant (C_8 |V(y)|+C_9)e^{-\vt |x|} \quad \text{for all} \quad (x,y)\in\R^2,
\end{equation}
where $C_8$, $C_9$, and $\vt>0$ are some fixed positive constants independent of $x$, $y$ and $\e$. In the space $L_2(\R^2)$ we consider a two-dimensional Schr\"odinger operator
\begin{equation}\label{A3}
\cH_\e=-\p_x^2-\p_y^2 + V_\e(x,y),\qquad V_\e(x,y):=V(y) + \e V_1(x,y)+\e^2 V_2(x,y)+\e^3 V_3(x,y,\e).
\end{equation}
The domain of this operator is the space $W_2^2(\R^2)\cap L_2(\R^2,(1+|V(y)|^2)\,d x \,d y)$. This operator is well-defined thanks to inequalities (\ref{A2}) and identity (\ref{A4}).

If the potential $V(y)$ is uniformly bounded, then the operator $\cH_\e$ is a particular case of a general operator studied in \cite{BZZ21}. In the case of an unbounded potential $V$, formally this is not true, but, nevertheless, all calculations and main results from \cite{BZZ21} can be easily extended for the considered operator $\cH_\e$. In order to do this, one just should replace the domain of the perturbed operator in \cite{BZZ21} by $W_2^2(\R^2)\cap L_2(\R^2,(1+|V(y)|^2)\,d x \,d y)$. Then the main results of \cite{BZZ21} adapted to the operator $\cH_\e$ read as follows.

The essential spectrum of the operator $\cH_\e$ is independent of $\e$ and is the half-line $[\L_1,+\infty)$. The numbers $\L_p$ are no longer eigenvalues both for the operators $\cH_\e$ and $-\p_x^2-\p_y^2+V(y)$ in $\R^2$, but become  thresholds in their essential spectra. For sufficiently small $\e$, these thresholds can bifurcate into eigenvalues located close to $\L_p$. Namely, fix $p\geqslant 1$ and denote:
\begin{equation}
K_{1,p} = -\frac{1}{2}\int\limits_{\R^2} \psi_p^2 V_1 \,dxd y, \qquad K_{2,p,\tau} = -\frac{1}{2}\int\limits_{\R^2}\left(\psi_p^2 V_2 - \psi_p V_1\cG_{p,\tau}( V_1 \psi_p)\right)\,dxd y,
\end{equation}
where  $\cG_{p,\tau}$ is an operator defined as
\begin{equation}\label{eq:Gp}
\begin{aligned}
\cG_{p,\tau}(f)(x,y) = \sum_{j=1}^{p-1} \frac{\iu\tau \psi_j(y)}{2\sqrt{\Lambda_p-\Lambda_j}}\int\limits_{\R^2}e^{i\tau\sqrt{\Lambda_p-\Lambda_j}|x-x_1|}\psi_j(y_1)f(x_1,y_1)\,d x_1 \,d y_1 \\
- \frac{1}{2}\psi_p(y)\int\limits_{\R^2}|x-x_1|\psi_p(y_1)f(x_1,y_1)\,d x_1\,d y_1 + U_p^\bot(x,y),
\end{aligned}
\end{equation}
where $U_p^\bot(x,y)\in W_2^2(\R^2)$ is the  unique  localized solution to  the inhomogeneous equation
\begin{equation}\label{eq:Ubot}
\big(-\pxx - \pyy +V(y) - \Lambda_p\big)U_p^\bot = f(x,y) - \sum_{j=1}^{p} f_j(x) \psi_j(y), \quad f_j(x) = \int\limits_{\R} f(x,y)\psi_j(y)\,d y,
\end{equation}
which satisfies the orthogonality conditions
\begin{equation}\label{eq:OrtCon}
\int\limits_{\R} U_p^\bot(x,y)\psi_j(y)\,d y=0, \qquad j=1,\ldots, p.
\end{equation}

We note that for $p=1$, the operator $\cG_{1,\tau}$ is independent of the choice of $\tau$ and the same is hence true for the constant $K_{2,1,\tau}$.

The main mathematical result is formulated in the following theorems which address the eigenvalues bifurcating from the edge (Theorem~\ref{thm:p=1}) and internal (Theorem~\ref{thm:p>1}) thresholds.

\begin{theorem} \label{thm:p=1}
	Let $p=1$. If $\RE\, (K_{1,1}+\e K_{2,1,\tau})>0$ for sufficiently small $\e$, that is, if
	\begin{equation*}
	\RE K_{1,1}>0 \qquad\quad \text{or}\quad\qquad  \RE K_{1,1}=0 \quad\text{and}\quad \RE K_{2,1,\tau}>0,
	\end{equation*}
	then the edge threshold $\L_1$ bifurcates into an eigenvalue  $\l_1(\e)$ of the operator $\cH_\e$. This eigenvalue is simple and has the asymptotic expansion
	\begin{equation*}
	\l_1(\e)=\L_1- \e^2\Big(K_{1,1}+\e K_{2,1,\tau }+O(\e^2)\Big)^2.
	\end{equation*}
\end{theorem}

\begin{theorem} \label{thm:p>1}
	Let $p\geqslant 2$. If $\RE\, (K_{1,{p}}+\e K_{2,{p}, {\tau}})>0$ and $\tau\IM\, (K_{1,{p}}+\e K_{2,{p},{\tau}})<0$ for some $\tau\in\{-1,+1\}$ and sufficiently small $\e$, that is, if
	\begin{equation*}
	\RE K_{1,{p}}>0 \qquad\quad \text{or} \quad\qquad  \RE K_{1,{p}}=0 \quad\text{and}\quad \RE K_{2,{p}, {\tau}}>0,
	\end{equation*}
	and
	\begin{equation*}
	\tau\IM K_{1,{p}}>0 \qquad\quad \text{or}\quad\qquad  \IM K_{1,{p}}=0 \quad\text{and}\quad \tau\IM K_{2,{p},\tau}<0,
	\end{equation*}
	then the internal threshold $\L_p$ bifurcates into an eigenvalue  $\l_{p,\tau}(\e)$ of the operator $\cH_\e$. This eigenvalue is simple and has the asymptotic expansion
	\begin{equation*}
	\l_{p,\tau}(\e)=\L_p- \e^2\Big(K_{1,p}+\e K_{2,p, {\tau}}+O(\e^2)\Big)^2.
	\end{equation*}
\end{theorem}

We stress that the edge threshold $\L_1$ can bifurcate only into a single simple eigenvalue, while internal thresholds $\L_p$, $p\geqslant 2$, can bifurcate into two simple eigenvalues. The latter situation occurs if the conditions of theorem~\ref{thm:p>1} are satisfied simultaneously for $\tau=-1$ and $\tau=1$.

\section{Calculation of the coefficient $ K_{2,p,\tau}$}
\label{sec:AppB}

Here we provide   auxiliary calculations which show how to obtain (\ref{eq:real}) and (\ref{eq:imag}) from  Eq.~(\ref{K2a}),
where  $U_{p,\tau}(x,y) :=  \cG_{p,\tau}\left(  \eta(x) V'(y)  \psi_p(y)\right)$.  This function is given  by the bulky expression which appears when $f=\eta(x) V'(y)  \psi_p(y)$ is substituted in (\ref{eq:Gp}). With a closer inspection,  we notice that the latter summand of the resulting expression, i.e., $U_p^\bot(x,y)$, is a real function.  Therefore it   only contributes to the real part of $K_{2,p,\tau}$. The middle summand is zero due to (\ref{eq:zero}). The main challenge is to handle the first summand, i.e., that in the form of   finite sum. For convenience, we redenote this term introducing a new function $W_{p,\tau}$.  Explicitly, we have
\begin{equation}\label{eq:W}
W_{p,\tau}(x,y) = \sum_{j=1}^{p-1} \frac{\iu\tau \psi_j(y)}{2\sqrt{\Lambda_p-\Lambda_j}} \int\limits_{\R^2}e^{i\tau\sqrt{\Lambda_p-\Lambda_j}|x-x_1|}\psi_j(y_1)\eta(x_1)V'(y_1)\psi_p(y_1)\,dx_1d y_1.
\end{equation}
Since $U_{p,\tau} = W_{p,\tau} + U^\bot_p$, we can rewrite (\ref{eq:real}) as
\begin{equation}
\label{eq:K2app}
K_{2,p,\tau}=\frac{1}{2}\int\limits_{\R^2}\left(\frac{\eta^2(x)}{2}V''(y)\psi_p^2(y) -\eta(x)V'(y)\psi_p(y)U_p^\bot(x,y)\right)\,dxdy + J_{p,\tau},
\end{equation}
where we introduced
\begin{equation}
\label{eq:J}
J_{p,\tau} := -\frac{1}{2}\int\limits_{\R^2}\eta(x)V'(y)\psi_p(y)W_{p,\tau}(x,y)\,dxd y.
\end{equation}
The main step is to compute  $J_{p,\tau}$. To this end, we introduce functions  $g_{p,j}(x) = e^{\iu \tau \sqrt{\Lambda_p-\Lambda_j}|x|}$ and treat each
$g_{p,j}(x)$ as a generalized function whose  Fourier transform reads  \cite{Brychkov}
\begin{equation}
\label{eq:hg}
\hg_{p,j}(s)=\iu\tau\left(\frac{1}{\sqrt{\Lambda_p-\Lambda_j}-s} - \frac{1}{-\sqrt{\Lambda_p-\Lambda_j}-s}\right) + \pi\big(\delta(s+\sqrt{\Lambda_p-\Lambda_j}) + \delta(s-\sqrt{\Lambda_p-\Lambda_j})\big),
\end{equation}
where $\delta$  denotes the Dirac delta function. According to  (\ref{eq:W}), the function $W_{p,\tau}$ contains the convolution of $g_{p,j}(x)$ and $\eta(x)$. Therefore, its Fourier transform can be computed as
\begin{equation}
\hat{W}_{p,\tau}(s,y) = \sum_{j=1}^{p-1}\frac{\iu\tau\cI_{p,j}\psi_j(y)}{2\sqrt{\Lambda_p-\Lambda_j}} \hg_{p,j}(s)\heta(s),
\end{equation}
where $ \cI_{p,j}$ are defined in (\ref{eq:cI}). Using Parseval's identity, from (\ref{eq:J}) we have
\begin{equation}
J_{p,\tau} = -\frac{1}{4\pi}\int\limits_{\R^2}\hat{W}_{p,\tau}(s,y)\heta(s)V'(y)\psi_p(y)\,d s\,d y = -\frac{\iu\tau}{8\pi}\sum_{j=1}^{p-1}\frac{\cI_{p,j}^2}{\sqrt{\Lambda_p-\Lambda_j}}\int\limits_{\R}\hg_{p,j}(s)|\heta(s)|^2\,d s.
\end{equation}

Using the explicit expression for Fourier transform $\hg_{p,j}$ in (\ref{eq:hg}) and also using that $|\heta(s)|^2$ is an even function of $s$, we compute
\begin{equation}
J_{p,\tau} =  -\frac{\iu\tau}{8\pi}\sum_{j=1}^{p-1}\frac{\cI_{p,j}^2}{\sqrt{\Lambda_{p}-\Lambda_{j}}}\left(
\frac{2\pi\iu}{\tau}
\cH_{p,j} + 2\pi\big|\heta(\sqrt{\Lambda_{p}-\Lambda_{j}})\big|^2
\right),
\end{equation}
where $\cH_{p,j}$ are Hilbert transforms defined in (\ref{eq:cH}).
The rest of   calculation is totally straightforward: we substitute $J_{p,\tau}$ back to (\ref{eq:K2app}), split it into real and imaginary parts and arrive at (\ref{eq:real}) and (\ref{eq:imag}).


\begin{thebibliography}{99}
	
\bibitem{Hatano1996}	N. Hatano and D. R. Nelson, Localization Transitions in Non-Hermitian Quantum Mechanics, Phys. Rev. Lett. {\bf 77}, 570 (1996).
	
\bibitem{Hatano1997} 	N. Hatano and D. R. Nelson, Vortex pinning and non-Hermitian quantum mechanics, Phys. Rev. B {\bf 56}, 8651 (1997).
	
	
\bibitem{Brouwer} 	P. W. Brouwer, P. G. Silvestrov, and C. W. J. Beenakker, Theory of directed localization in one dimension, Phys. Rev. B {\bf 56}, R4333(R) (1997).

\bibitem{Efetov} K. B. Efetov, Directed Quantum Chaos, Phys. Rev. Lett.  {\bf  79}, 491  (1997).

	
\bibitem{Goldsheid}	I. Ya. Goldsheid and B. A. Khoruzhenko, Distribution of Eigenvalues in Non-Hermitian Anderson Models, Phys. Rev. Lett. {\bf 80}, 2897 (1998).

\bibitem{Hatano1998}		N. Hatano and D. R. Nelson, Non-Hermitian delocalization and eigenfunctions, Phys. Rev. B {\bf 58}, 8384 (1998).
	
\bibitem{Yurkevich}	I. V. Yurkevich and I. V. Lerner, Delocalization in an Open One-Dimensional Chain in an Imaginary Vector Potential, Phys. Rev. Lett. {\bf 82}, 5080 (1999).
	
	
\bibitem{Takeda} 	K. Takeda and I. Ichinose, Random-Mass Dirac Fermions in an Imaginary Vector Potential: Delocalization Transition and Localization Length, J. Phys. Soc. Jap. {\bf 70}, 3623 (2001).
	
\bibitem{CompGauge2D}	T. Kuwae and N. Taniguchi, Two-dimensional non-Hermitian delocalization transition as a probe for the localization length, Phys. Rev. B \textbf{64}, 201321(R) (2001).
	
\bibitem{Heinrichs} 	J. Heinrichs, Eigenvalues in the non-Hermitian Anderson model, Phys. Rev. B {\bf 63}, 165108 (2001).

\bibitem{analogy} S. Longhi, Quantum-optical analogies using photonic structures, Laser \& Photon. Rev. {\bf  3},   243--261 (2009).
			
\bibitem{LonghiGattiPRB} 	S. Longhi, D. Gatti, and G. D. Valle, Non-Hermitian transparency and one-way transport in low-dimensional lattices by an imaginary gauge field, Phys. Rev. B {\bf 92}, 094204 (2015).
	
\bibitem{LonghiGattiSciRep} 	S. Longhi, D. Gatti, and G. D. Valle, Robust light transport in non-Hermitian photonic lattices, Sci. Rep. {\bf 5}, 13376 (2015).
	
\bibitem{Longhi2017PRA} 	S. Longhi,  Nonadiabatic robust excitation transfer assisted by an imaginary gauge field, Phys. Rev. A {\bf 95}, 062122 (2017).
	
	
\bibitem{Qin} Chengzhi Qin, Bing Wang,  Zi Jing Wong, S.  Longhi,  and Peixiang Lu, Discrete diffraction and Bloch oscillations in non-Hermitian frequency lattices induced	by complex photonic gauge fields, Phys. Rev. B {\bf  101}, 064303 (2020).
	
\bibitem{Zhang} Lei Du, Yan Zhang and Jin-Hui Wu, Controllable unidirectional transport and light trapping using a one-dimensional lattice with non-Hermitian coupling, Sci. Rep. {\bf 10}, 1113  (2020).
	
	
	
	
	
\bibitem{Lana} 	 L. Descheemaeker, V. Ginis, S. Viaene, and P. Tassin, Optical Force Enhancement Using an Imaginary Vector Potential for Photons, Phys. Rev. Lett. {\bf 119}, 137402 (2017).
	
	
\bibitem{GaugeReview2} J. Dalibard,  F. Gerbier, G. Juzeli\~unas, and P.  \"Ohberg,  Colloquium: Artificial gauge potentials for neutral atoms, Rev. Mod. Phys.  {\bf 83}, 1523--1543 (2011).	 	

\bibitem{Galitski} V. Galitski and  I. B. Spielman, Spin–orbit coupling in quantum gases, Nature {\bf 494}, 49–54 (2013).
	
\bibitem{GaugeReview}	 N. Goldman, G. Juzeliunas, P. \"Ohberg, and I. B. Spielman, Light-induced gauge fields for ultracold atoms, Rep. Prog. Phys. {\bf 77}, 126401 (2014).

	
\bibitem{polaritons} H. T. Lim, E.   Togan, M.  Kroner, J.   Miguel-Sanchez, and  A.  Imamoğlu, Electrically tunable artificial gauge potential for polaritons, Nat.   Commun. {\bf  8}, 14540 (2017). 
	
\bibitem{optics} Y. Lumer, M. A. Bandres, M. Heinrich, L. J. Maczewsky, H. Herzig-Sheinfux, A. Szameit, and   M.  Segev, Light guiding by artificial gauge fields, Nat. Photonics  {\bf  13}, 339--345 (2019). 
	
\bibitem{circuits}  Jiexiong Wu, Zhu Wang, Yuanchuan Biao, Fucong Fei, Shuai Zhang, Zepeng Yin, Yejian Hu, Ziyin Song, Tianyu Wu, Fengqi Song, and   Rui Yu, Non-Abelian gauge fields in circuit systems,  Nat. Electron {\bf  5}, 635--642 (2022).  
	 	
\bibitem{Zez21} D. A. Zezyulin,  Y.  V. Kartashov,  and V.  V. Konotop, Superexponential amplification, power blowup, and solitons sustained 	by non-Hermitian gauge potentials, Phys. Rev. A {\bf  104}, L051502 (2021).
	
\bibitem{PPT} J. Yang, Partially $\PT$ symmetric optical potentials 	with all-real spectra and soliton families in multidimensions, Opt. Lett. {\bf 39}, 1133--1136 (2014).
	
\bibitem{Znojil} M. Znojil, $\PT$-symmetric harmonic oscillators, Phys. Lett. A {\bf  259}, 220 (1999).
	
\bibitem{Bender} C. M. Bender and H. F. Jones, Interactions of Hermitian and non-Hermitian
Hamiltonians, J. Phys. A: Math. Theor {\bf  41}, 244006 (2008).
	
\bibitem{KSZ12} V. V. Konotop, V. S. Shchesnovich, D.  A. Zezyulin, Giant amplification of modes in parity-time symmetric waveguides, Phys. Lett. A {\bf 376},  2750--2753 (2012).
	
\bibitem{ZezKon12}  D. A. Zezyulin and V. V. Konotop, Nonlinear modes in the harmonic $\PT$-symmetric potential, Phys. Rev. A   {\bf 85}, 043840 (2012).
	
\bibitem{Gallo}  C. Gallo and D. Pelinovsky, On the Thomas–Fermi Approximation of the Ground State in a $\PT$-Symmetric Confining Potential, Stud. Appl. Math. {\bf 133}, 398--421 (2014).
	
\bibitem{Garmon} 	S. Garmon, M. Gianfreda, and N. Hatano, Bound states, scattering states, 	and resonant states in $\PT$-symmetric open quantum systems, 	{ Phys. Rev. A} {\bf 92},  022125 (2015).
	
\bibitem{Yang17} J. Yang, Classes of non-parity-time-symmetric optical potentials with exceptional-point-free phase transitions, {Opt. Lett.} {\bf 42},  4067--4070 	(2017).
	
\bibitem{KZ17}  V. V. Konotop and D. A. Zezyulin, Phase transition through the	splitting of self-dual spectral singularity in optical potentials, 	{Opt. Lett.} {\bf 42},  5206--5209   (2017).
	
	
\bibitem{Most} A. Mostafazadeh, Self-dual Spectral Singularities and Coherent Perfect Absorbing Lasers without $\PT$-symmetry, J. Phys. A: Math. Theor. {\bf  45}, 444024 (2012).
	
\bibitem{Longhi} S. Longhi,  $\PT$-symmetric laser absorber, Phys. Rev. A {\bf  82}, 031801(R) (2010).
	
	
\bibitem{BICoptics} 	S. Longhi, Bound states in the continuum in $\PT$-symmetric 	optical lattices, {  Opt. Lett.} {\bf 39},  1697--1700  (2014).
	
	
\bibitem{Kartashov} 	Y. V. Kartashov, C. Mili\'an, 	V. V. Konotop, and L. Torner, Bound states in the continuum 	in a two-dimensional PT-symmetric system, { Opt. Lett.} {\bf 43}, 575--578  (2018).
	
	
\bibitem{BZZ21}  D. I. Borisov, D. A. Zezyulin, and M. Znojil, Bifurcations of thresholds in essential spectra of	elliptic operators under localized	non-Hermitian perturbations, Stud. Appl. Math. {\bf 146}, 834--880 (2021).
	
\bibitem{Brychkov}  Yu. A. Brychkov and A. P. Prudnikov,  Integral transforms of generalized functions, Gordon and Breach Sci. Publ. New-York. 1989.
	
\bibitem{razavy1980} 	M. Razavy, An exactly soluble Schr\"odinger equation with a bistable potential, Am. J. Phys. \textbf{48}, 285 (1980).

\bibitem{EPsolitons} D. A. Zezyulin,  Y.   V. Kartashov,  and V.   V. Konotop, Metastable two-component solitons near an exceptional point, Phys. Rev. A {\bf   104}, 023504 (2021).

\bibitem{Metafune} G. Metafune and  R. Schnaubelt, The domain of the Schr\"odinger operator $-\Delta+x^2y^2$, Note Mat. {\bf 25}, 97--103  (2005/06).

\bibitem{EvGi} W. N. Everitt and  M. Giertz. Inequalities and separation for certain ordinary differential operators,   Proc. London Math. Soc. (3)  {\bf  28},  352--372  (1974).


\bibitem{Lieb} E. H. Lieb and M. Loss, Analysis -- 2nd ed.   AMS. Providence. 2001.

\bibitem{Kato} T. Kato, Perturbation theory for linear operators, Springer-Verlag, Berlin. 1966.








\end{thebibliography}
\end{document}